\documentclass[12pt]{iopart}
\usepackage{iopams}
\usepackage{graphicx}
\usepackage{color}

\begin{document}
\title{Experimental study of hard X-rays emitted from meter-scale positive discharges in air}
\author{P O Kochkin$^1$, C V Nguyen$^1$, A P J van Deursen$^1$ and U Ebert$^2$}
\address{$^1$ Department of Electrical Engineering, Eindhoven University of Technology, POBox.~513, NL-5600~MB Eindhoven, The Netherlands}
\address{$^2$ Department of Applied Physics, Eindhoven University of Technology, and Centre for Mathematics and Computer Science (CWI), POBox~94079, NL-1090~GB Amsterdam, The Netherlands}
\ead{p.kochkin@tue.nl, a.p.j.v.deursen@tue.nl, ute.ebert@cwi.nl}

\begin{abstract}

We investigate structure and evolution of long positive spark breakdown; and we study at which stage pulses of hard X-rays are emitted. Positive high-voltage pulses of standardized lightning impulse wave form of about 1~MV were applied to about 1~meter of ambient air. The discharge evolution was imaged with a resolution of tens of nanoseconds with an intensified CCD camera. LaBr$_{3}$(Ce$^+$) scintillation detectors recorded the X-rays emitted during the process. The voltage and the currents on both electrodes were measured synchronously. All measurements indicate that first a large and dense corona of positive streamers emerges from the high voltage electrode. When they approach the grounded electrode, negative counter-streamers emerge there, and the emission of hard X-rays coincides with the connection of the positive streamers with the negative counter-streamers. Leaders are seen to form only at later stages.

\vspace{2 cm}{Please cite as P O Kochkin et al 2012 J. Phys. D: Appl. Phys. 45 425202 doi:10.1088/0022-3727/45/42/425202}
\end{abstract}

\maketitle


\section{Introduction}
\label{sec:introduction}

The generation of X-rays from electrical discharges
at near to atmospheric pressure requires that electrons reach very high energies despite their numerous collisions with gas molecules; the X-rays are then generated by Bremsstrahlung. Wilson \cite{Wilson1925} suggested that favourable conditions for electron run-away to high energies could exist in thunderclouds. After unsuccessful attempts in the 1930ies it took many years before the association of X-rays and thunderstorms could be established. In Chapters 2 and 6 of his book \cite{Babich2003} Babich presents an extensive overview of the early and later experiments. McCarthy \cite{McCarthy1985} observed X-rays of the order of 10 keV in aircraft flights through thunderstorms in the 1980ies. At surface level, X-rays associated with the approaching lightning leader have been detected in triggered lightning measurements \cite{Dwyer2003, Dwyer2004, Dwyer2011}; their energy is of the order of 250 keV. Terrestrial gamma-ray flashes from thunderstorms with energies of up to 20 MeV \cite{Smith2005} or even up to 40 MeV \cite{Marisaldi2010} are measured from satellites.

X-rays can also occur during the initiation of sparks in the laboratory on a scale of millimetres up to meters, even with standard atmospheric conditions of approximately 1~bar and 20~$^{\circ}$C. Unrolling adhesive tapes generates X-rays \cite{Camara2008}, as do helium filled spark chambers \cite{Frankel1966}, the formation of corona discharges \cite{Nguyen2010} on a centimetre scale, or meter long discharges \cite{Dwyer2008, Nguyen2008, March2011, Cooray2009}. X-rays are a common feature of a high-voltage discharge, regardless of size.

While in short gaps the accelerated electrons actually might emit their Bremsstrahlung during collision with surrounding metal equipment, experiments in sufficiently long air gaps are actually similar to thunderstorm conditions as the Bremsstrahlung photons are generated by collisions with air molecules as well. We will argue that for the experiments reported here the photons are indeed generated in open air.

As shown in recent simulations \cite{Moss2006,Li2009,Chanrion2010,Celestin2011}, the highly enhanced electric field at the tip of a negative streamer discharge could actually accelerate the electrons into the run-away regime, and the effect is even stronger at the tip of a naked lightning leader with its larger dimensions and higher voltage \cite{Xu2012} if such a naked corona-less leader exists. We will show below how the streamers develop in our experiment, and at which stage they actually emit the hard X-rays. The effect is created here not by a single streamer, but by the complete streamer corona.

We study the initiation phase before breakdown. Such studies have been performed earlier by several groups. Here we refer in particular to the early work in France \cite{Reess1995} \cite{LesRenardieres1977}. Les Renardieres group investigated static, streak and Schlieren photos of positive discharges. They described first corona formation followed by leader propagation, final jump and return stroke. But they did not measure X-rays emitted from a spark during this processes. We measure the voltage and the current on both electrodes, synchronized with nanosecond fast photography and X-ray detection. Our aim is to find where and when the X-rays are generated.

\section{Experimental setup}
\label{sec:experimental_setup}

The 2~MV Marx generator at Eindhoven University of Technology delivers an IEC standardized lightning impulse voltage waveform of 1.2/50~$\mu$s rise/fall time when not loaded. The generator is connected to a spark gap with cone-shaped electrodes. The setup is similar to the one described in \cite{Nguyen2008}, with five changes worth mentioning here. First, the discharge channel is vertical rather than horizontal. This provided a more symmetrical environment of the discharge than in \cite{Nguyen2008}. Second, conductive Velostat plastic covered all sharp protruding objects close to the gap, in order to avoid local streamer formation. Third, current probes were mounted on both electrodes. Fourth, an intensified CCD camera was used. Finally, two scintillator detectors were placed at varying positions. The goal was to find the angular distribution of the X-ray emission.

The distance between the high-voltage (HV) and grounded (GND) electrodes was 1 meter. The applied voltage between electrodes was about 1~MV and did not change within all experimental study.

\begin{figure*}[t]
\centering
\includegraphics[width=150mm]{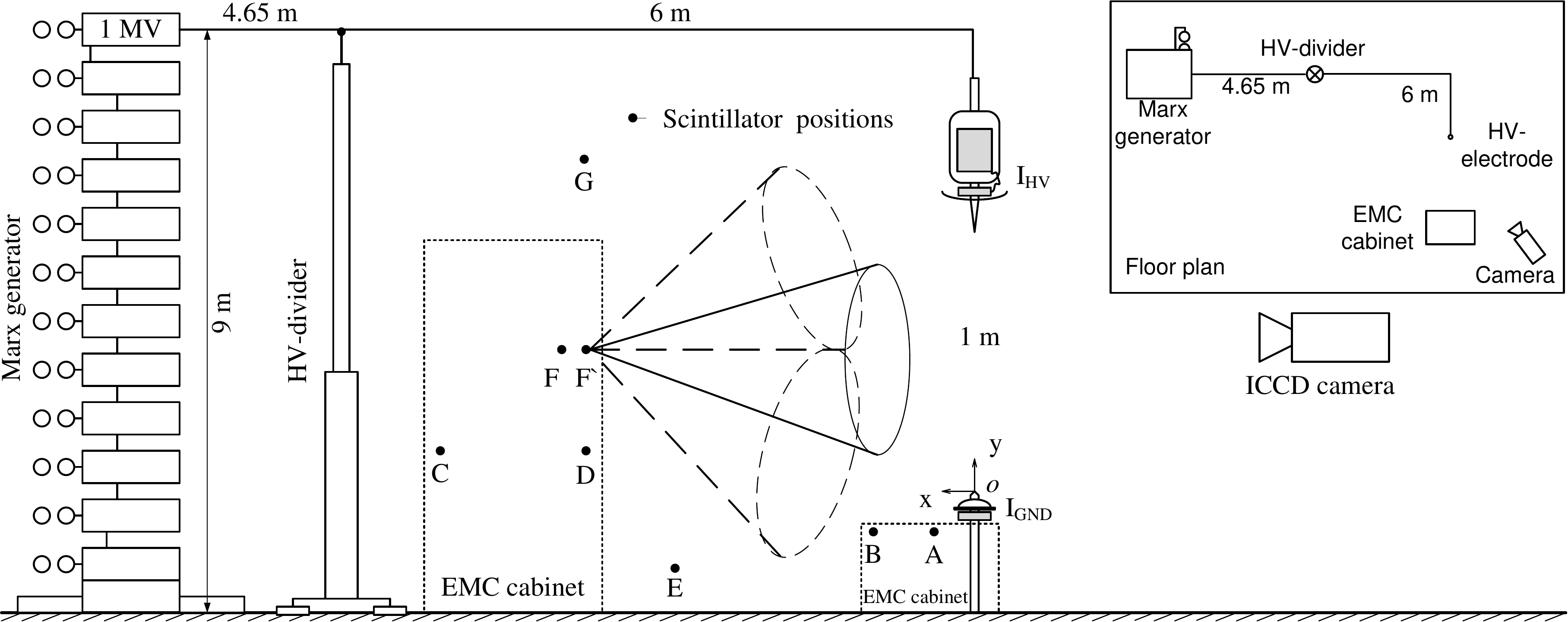}
\caption{Schematic sketch of the spark-gap geometry and positions of X-ray detectors. Each dot labelled from A up to G corresponds to a LaBr$ _{3} $(Ce$ ^{+} $) detector scintillator position; all positions are in a single vertical plane. Cones indicate the field of view of the detector when placed inside the lead collimator discussed in \Sref{sec:the_region_of_xray_emission}. The ICCD camera is located at 4~m distance from the gap; it is shown in correct relative height, but not in the scaled position. The distance between Marx generator and the spark gap is 8~m. The upper right inset shows the correctly scaled floor plan of the setup.}
\label{fig:setup}
\end{figure*}

To obtain time-integrated and time-resolved images of the pre-breakdown phenomena, a ns-fast 4Picos ICCD camera was located at 4~m distance from the gap, perpendicular to discharge-developing axes. The camera was placed inside a small but well-protected cabinet. The cables between the camera cabinet and the large cabinet with registration equipment (\Fref{fig:setup}) were about 5~m long. All cables were placed inside a stainless steel flexible hose. The hose and its mounting were near to vacuum tight; this ensured adequate electromagnetic compatibility (EMC) and interference-free operation of the camera. The field of view of the camera just covered the full gap, including grounded and high-voltage electrodes. The optical gain by a built-in image intensifier was 10$^{4}$. The camera has a black and white CCD; it was not calibrated. However, the camera parameters and optical system remained the same, except only one parameter: the exposure time.

Two LaBr$ _{3} $(Ce$ ^{+} $) scintillator detectors manufactured by Saint-Gobain were mounted in EMC cabinets and recorded the X-rays. Any interference on their signals due to the discharge initiation can be excluded, since such interference would most likely manifest itself as oscillatory signal, and not mimic a clear scintillator signal. Also, in many discharges only the signal channel noise floor was measured. The scintillators have a fast primary rise/decay time (11/16~ns) and a high light yield of 63~photons/keV, which is 165\% of the more common NaI(TI). The linearity of the detectors has been tested \cite{NguyenCV2012} on $ ^{241} $Am, $ ^{137} $Cs, $ ^{60} $Co and remains perfect up to 2505 keV, which is the total absorbed energy from two gamma quanta of the $ ^{60} $Co source in the scintillator. The slight deviation from linearity at higher energies is attributed to saturation of the photomultiplier. The output of the photomultiplier is recorded directly on the oscilloscope without any waveshaping electronics usually employed in photon counting. This allows to distinguish individual pulses even when pile-up occurs within the decay time of the scintillator. In some series of measurements the detectors were located inside separate EMC-cabinets; one detector was placed in a small cabinet at positions A and B and the other detector was placed in the large cabinet at positions C, D, F, F' as indicated in \Fref{fig:setup}.

The total aluminium thickness between the
scintillator and surrounding area is 550~$ \mu $m. This shielding is almost transparent for photons with an energy above 30~keV (attenuation 15\% or less). To determine the origin and the energy composition of the X-ray signals, we used up to 2.5~cm thick lead
attenuators and collimators around one detector. The attenuators effectively absorb photons with energies up to 1~MeV (10 times attenuation). Attenuation by 2~m of air between the spark gap and detector is 8\% including scattering and photoelectric absorption for 30~keV photons. At 200~keV energy level the air impact is less than 3\%. We did not take attenuation by air into account.

Some X-ray signals corresponded to a total absorbed energy of about 3.4~MeV. We determined the attenuation curve by measurements with lead attenuators of different thickness. The slope of the curves contains information about the incident radiation. Additional measurements with the 662~keV gammas of $ ^{137} $Cs confirmed the analysis (see \Sref{sec:energy_spectra}).

The spark gap with two cone-shaped electrodes deviates from the more commonly used point-plane configuration. Two cones provide clear points for the corona discharge to start, and allow more precise measurements of currents due to moving charges. Two Pearson 7427 current probes detected the pre-breakdown currents on both electrodes. In order to protect the probes against a direct discharge we mounted a vaulted aluminium disk between probe and electrode tip. The probe for the HV electrode had an optical transmission system. Suitable attenuators and two antiparallel high speed diodes protected the input of the transmitter. The diodes limited the linear response to 250~A. The GND electrode and the probe were mounted on top of the small EMC cabinet containing the scintillator detector (\Fref{fig:setup}).

The Marx generator HV divider (80.000:1) was used for voltage measurements. In spite of the large distance between divider and gap, this voltage measurement sufficed to distinguish the different phases of the discharge development. That voltage recording should not be considered as the gap voltage, because of the inductive effects caused by the 8~m long leads between divider top and spark gap HV electrode. Two 4-channel Lecroy oscilloscopes with a maximum sampling rate of 5~Giga-Samles/s collected the data of the X-rays, current and voltage synchronously.

\section{Results}
\label{sec:results}

Several reports on the observation of X-rays with lightning impulse generators have been published \cite{Dwyer2008,Nguyen2008,March2011}. Up to a few meter distance between detector and discharge is usually maintained because of safety. In this study X-rays are observed in many, but not in all discharges. We studied the formation of X-rays during the initiation of 951 discharges of positive polarity. The minimal time from one discharge to the next was 10 seconds. In some discharges, we observed up to three easily recognizable individual X-ray peaks with different times and amplitudes. This allowed us to obtain statistical information on the energy spectra and the time distribution averaged over many HV discharges.

\subsection{Discharge development process}
\label{subsec:discharge_development_process}

\begin{figure}[ht]
\begin{minipage}[b]{0.18\linewidth}
\includegraphics[width=30mm]{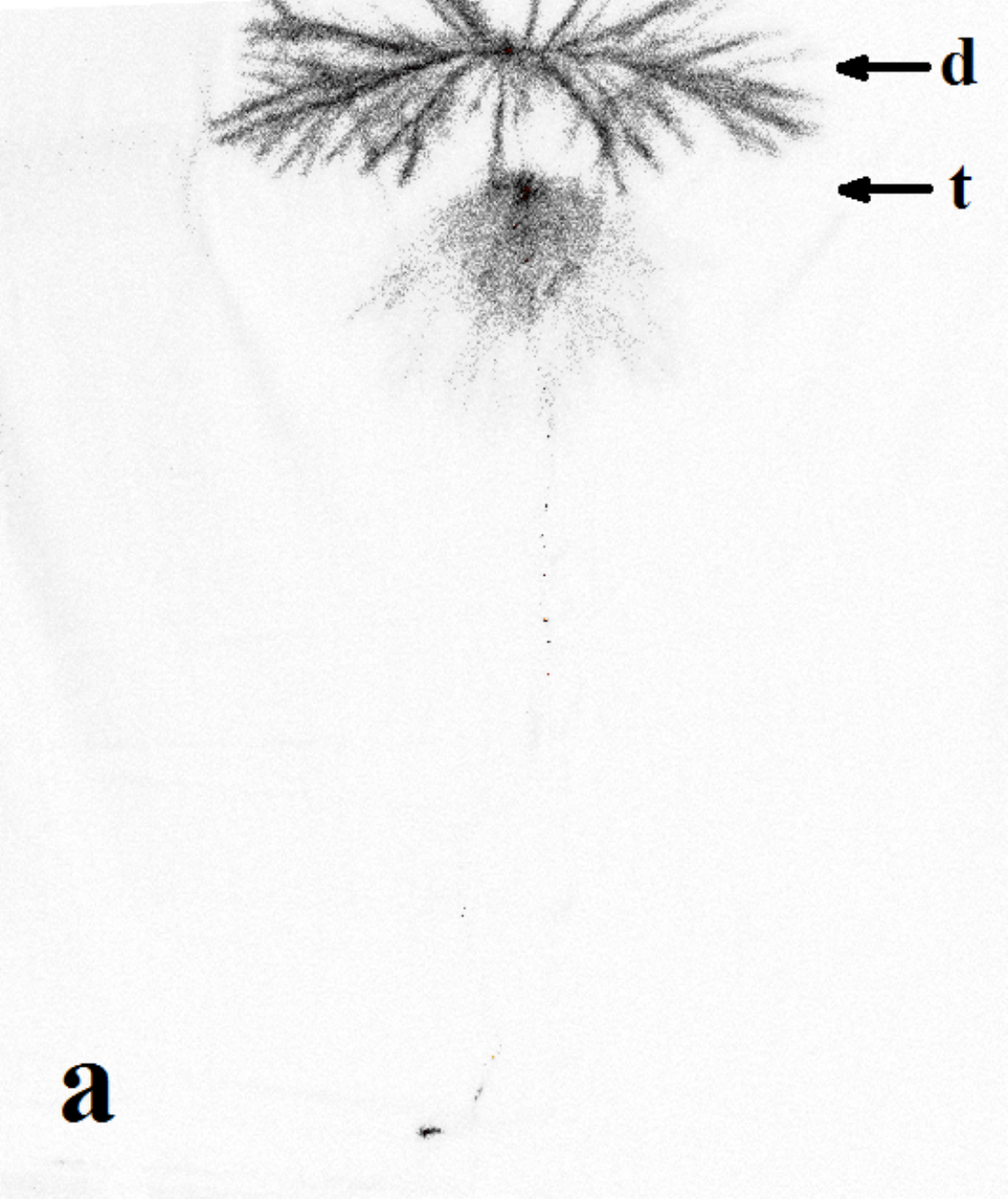}\end{minipage}
\begin{minipage}[b]{0.18\linewidth}
\includegraphics[width=30mm]{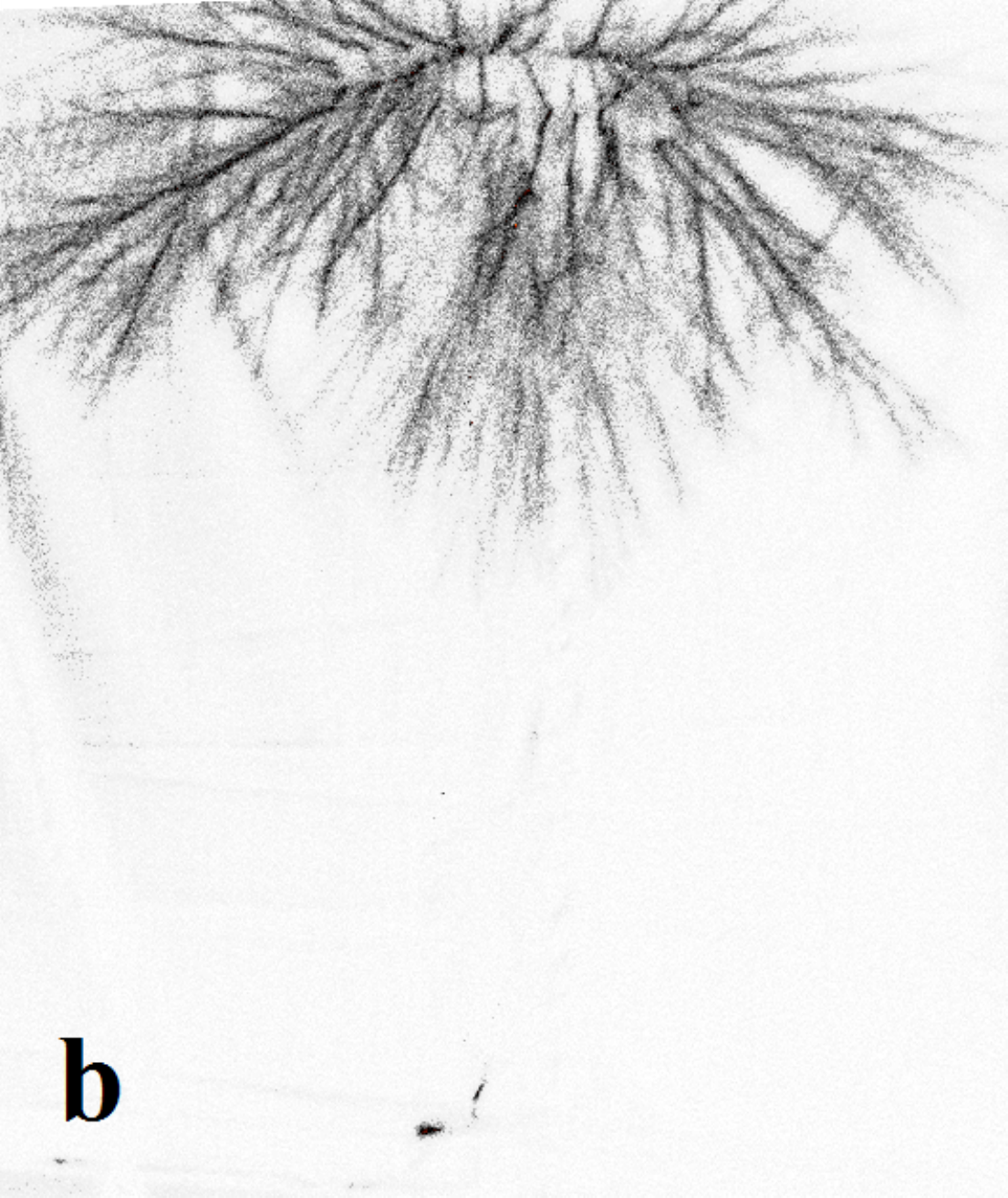}\end{minipage}
\begin{minipage}[b]{0.18\linewidth}
\includegraphics[width=30mm]{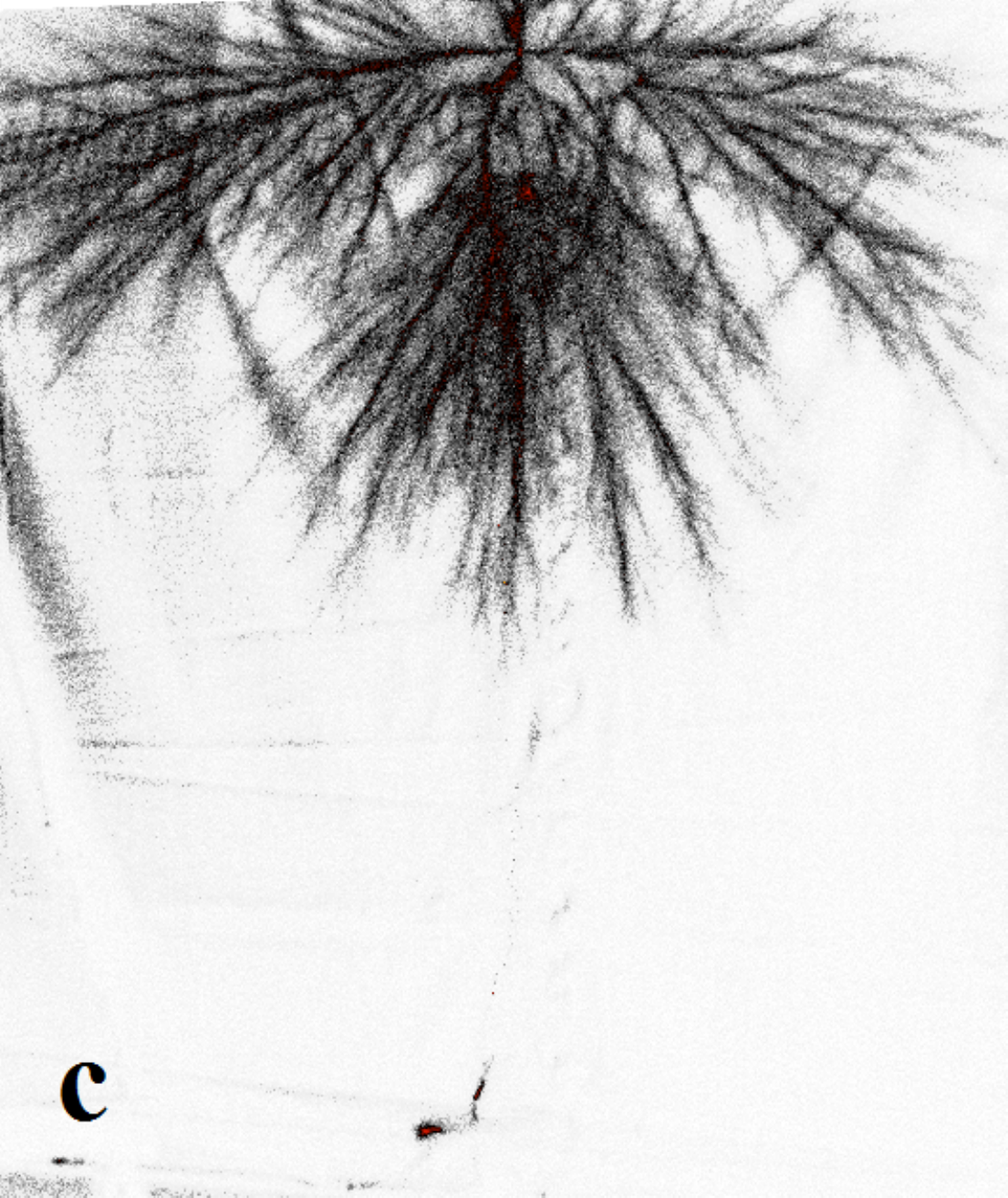}\end{minipage}
\begin{minipage}[b]{0.18\linewidth}
\includegraphics[width=30mm]{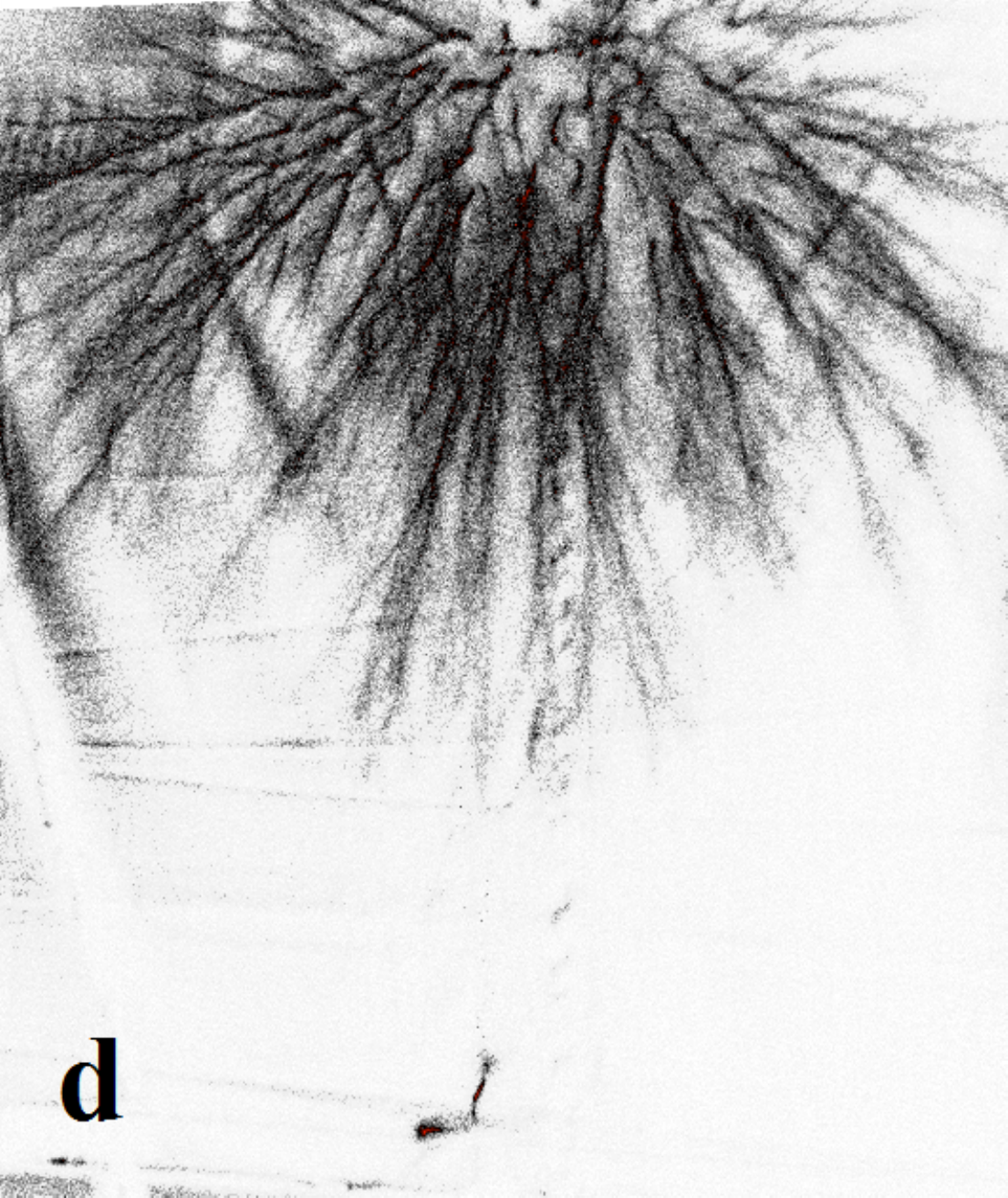}\end{minipage}
\begin{minipage}[b]{0.18\linewidth}
\includegraphics[width=30mm]{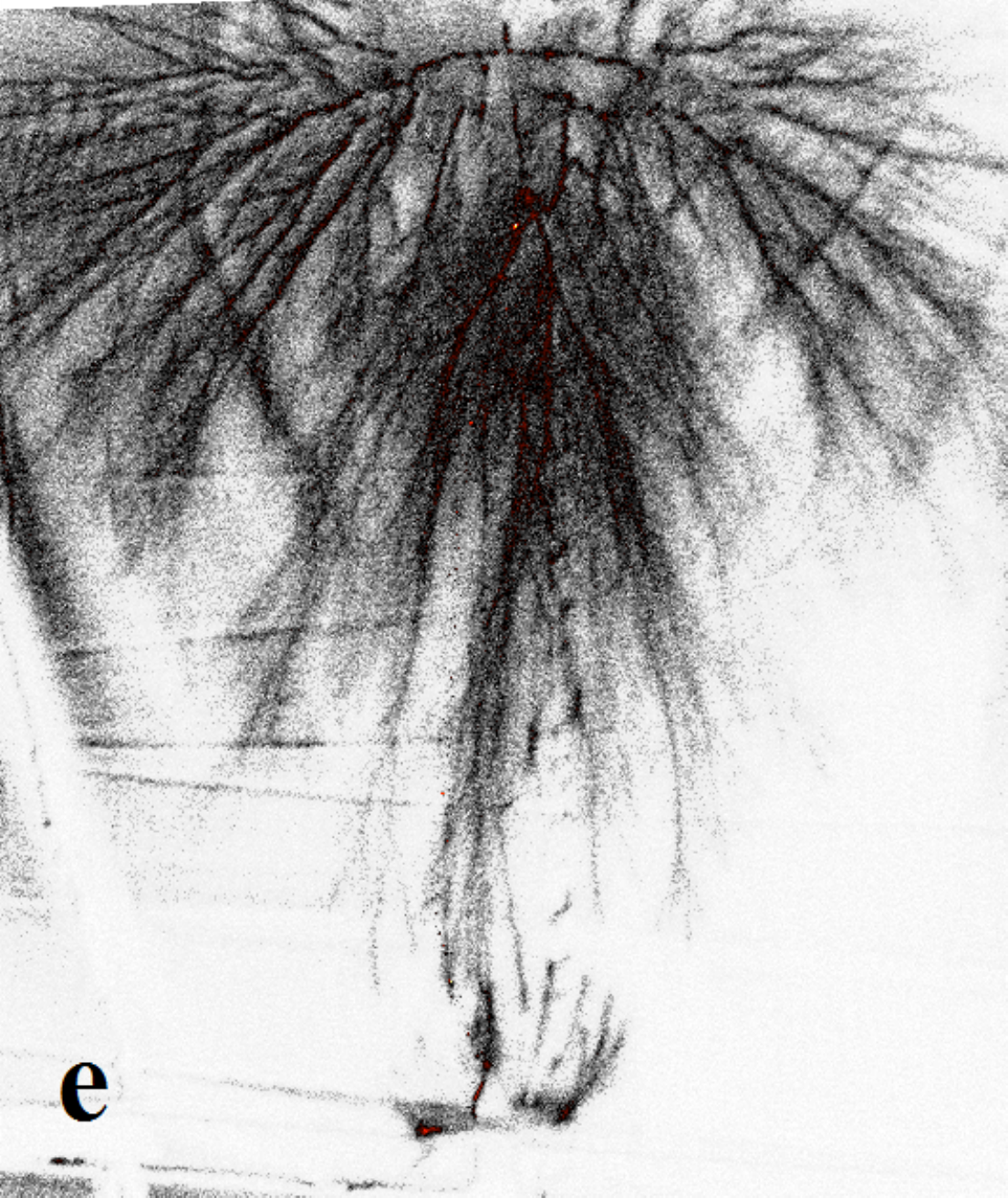}\end{minipage}
\begin{minipage}[b]{0.18\linewidth}
\includegraphics[width=30mm]{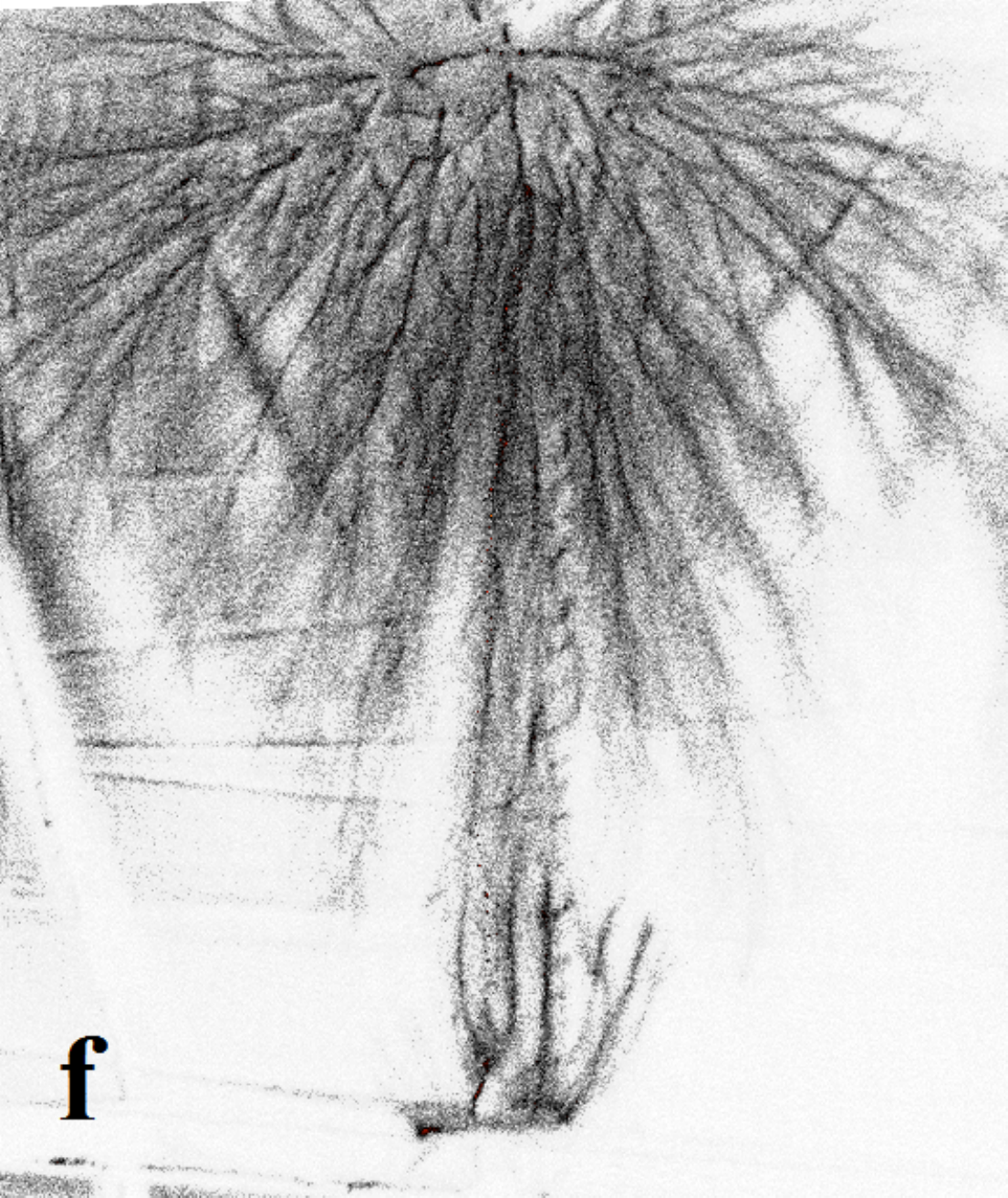}\end{minipage}
\begin{minipage}[b]{0.18\linewidth}
\includegraphics[width=30mm]{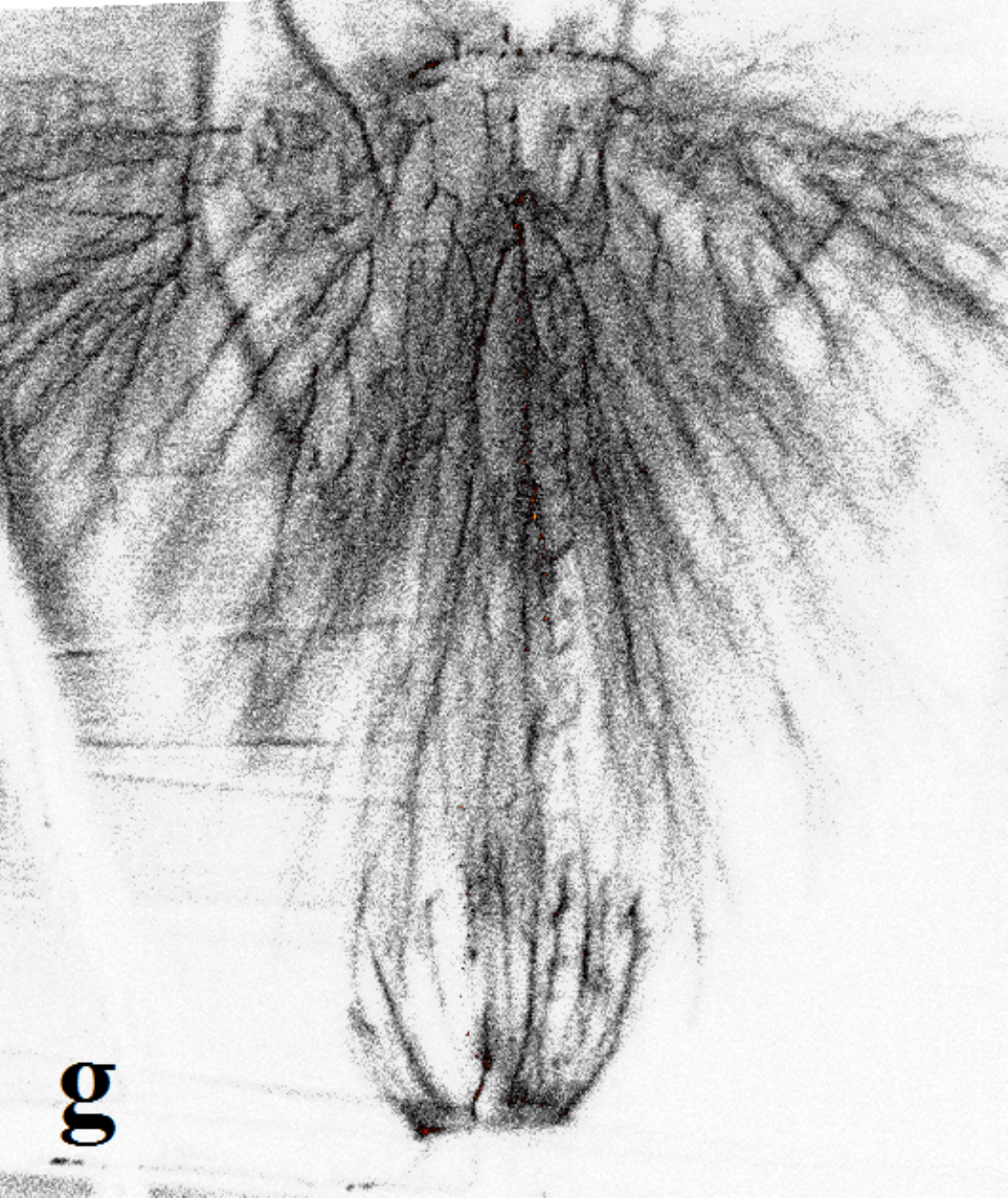}\end{minipage}
\begin{minipage}[b]{0.18\linewidth}
\includegraphics[width=30mm]{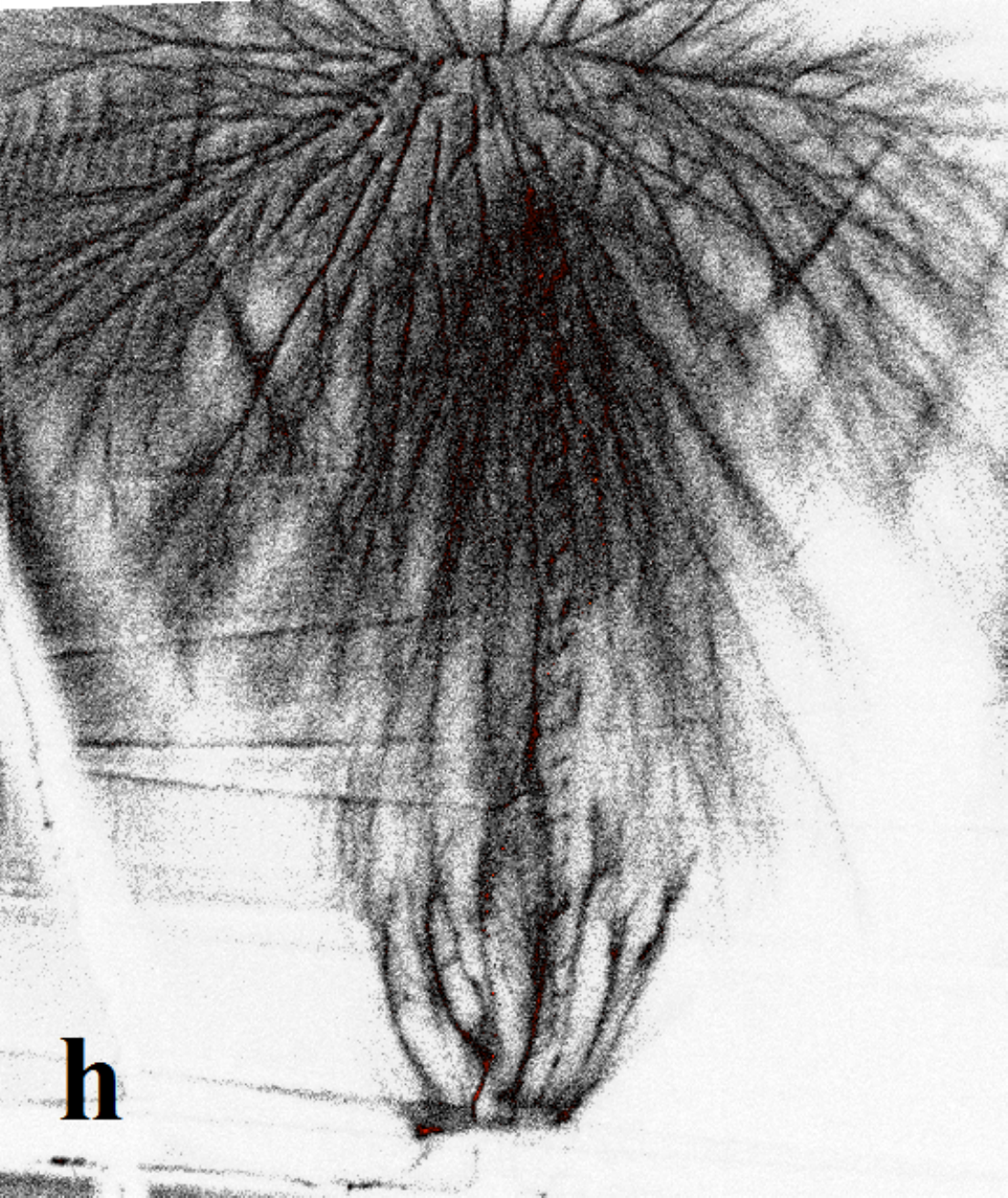}\end{minipage}
\begin{minipage}[b]{0.18\linewidth}
\includegraphics[width=30mm]{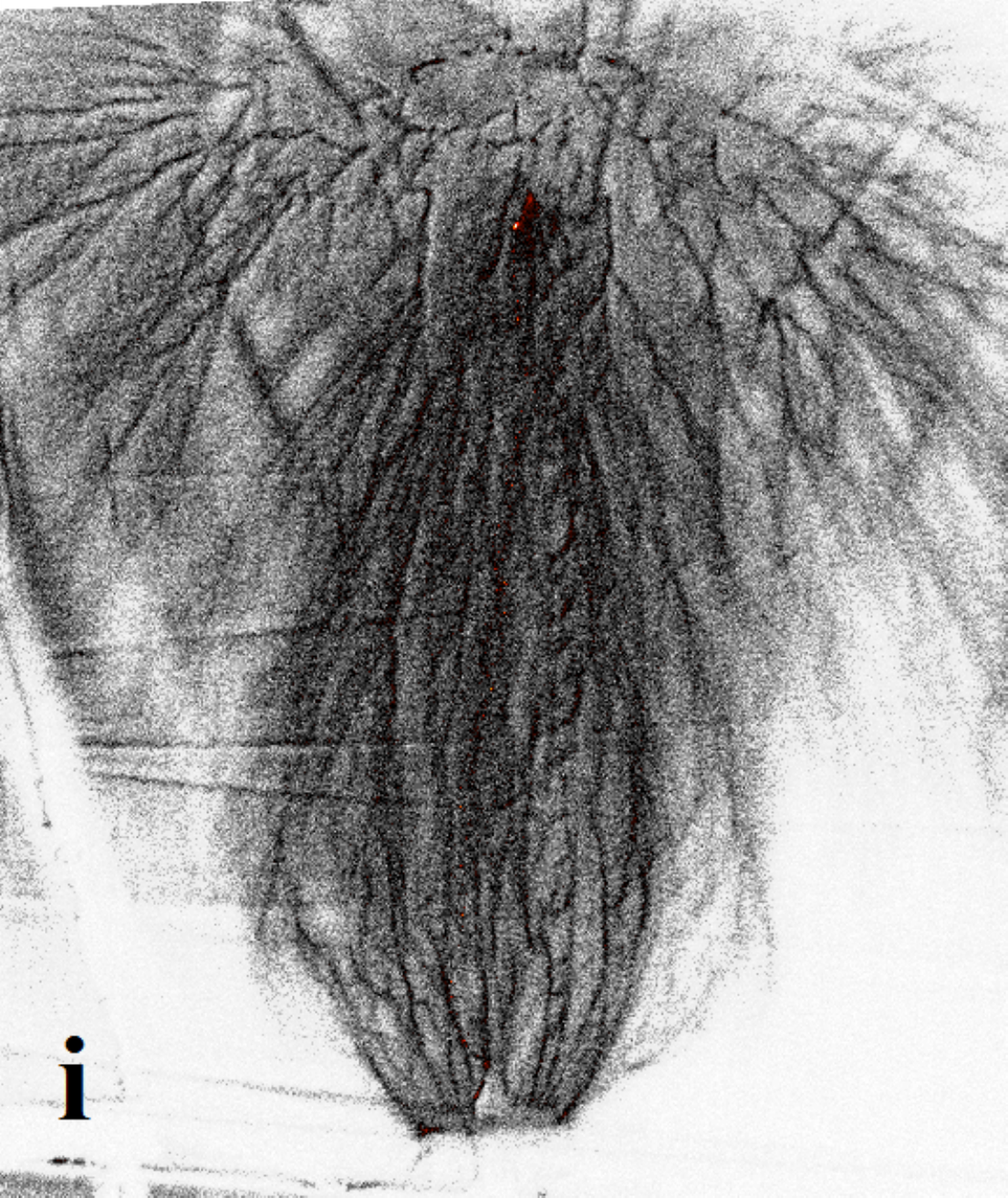}\end{minipage}
\begin{minipage}[b]{0.18\linewidth}
\includegraphics[width=30mm]{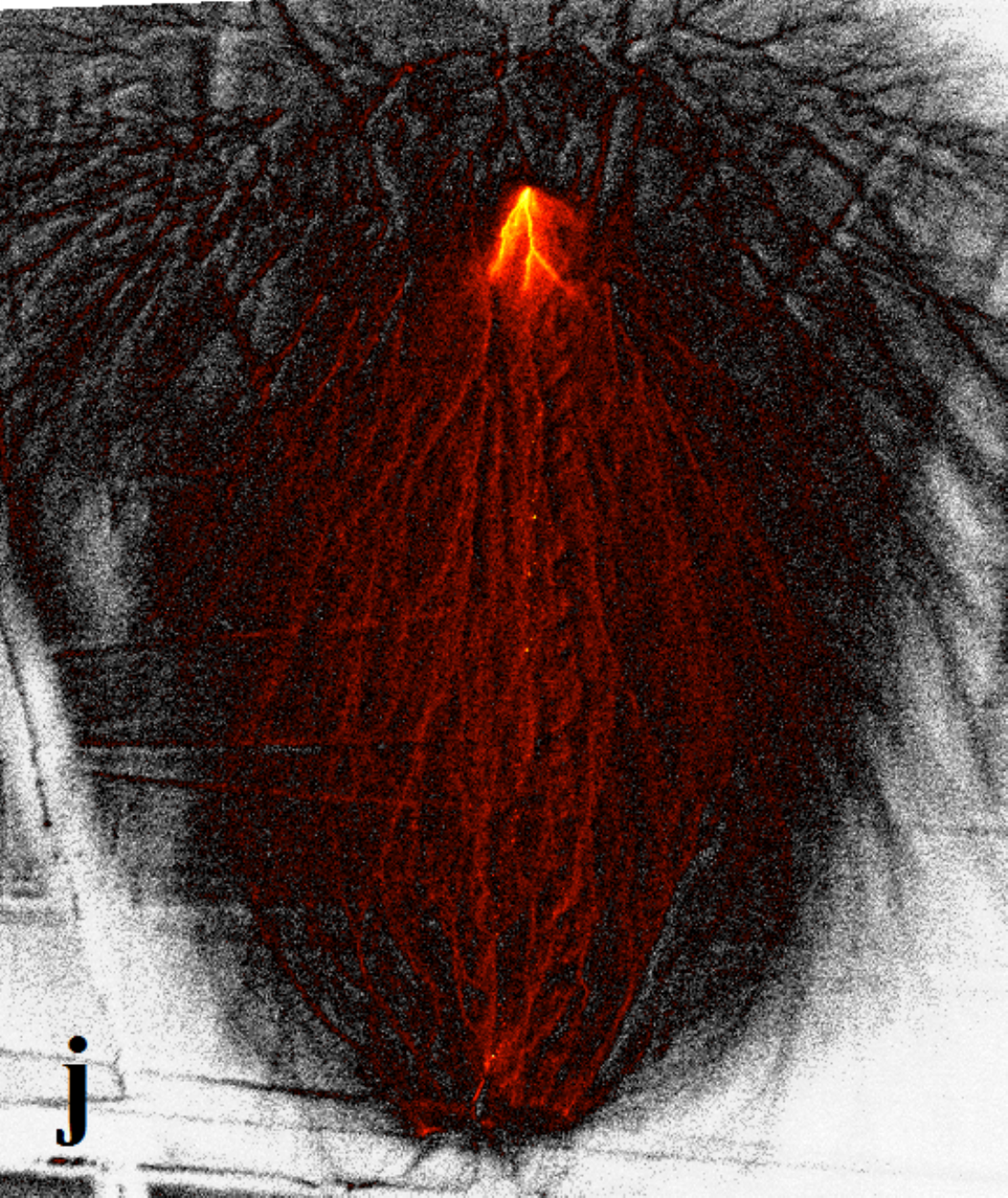}\end{minipage}
\begin{minipage}[b]{0.18\linewidth}
\includegraphics[width=30mm]{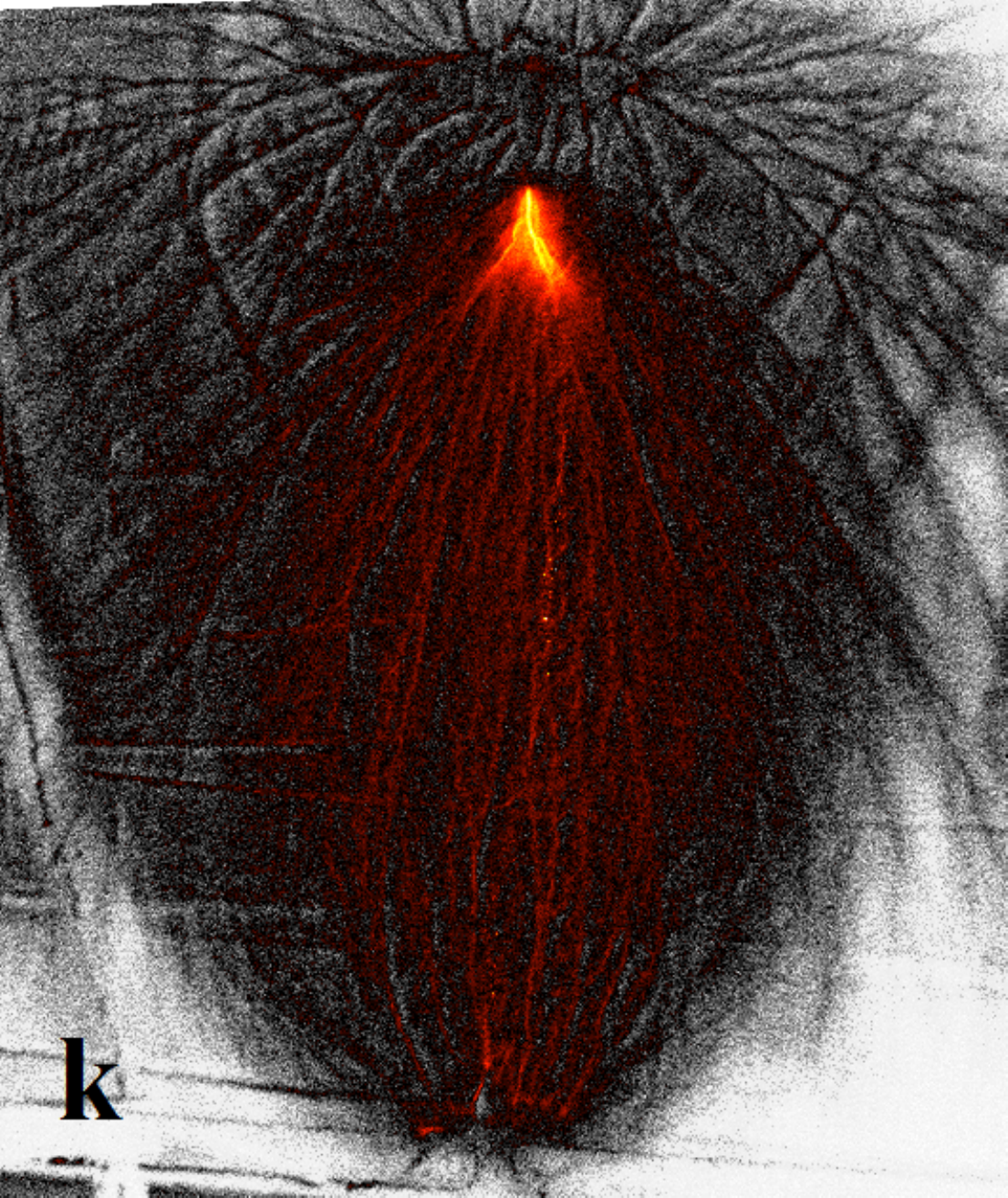}\end{minipage}
\begin{minipage}[b]{0.18\linewidth}
\includegraphics[width=30mm]{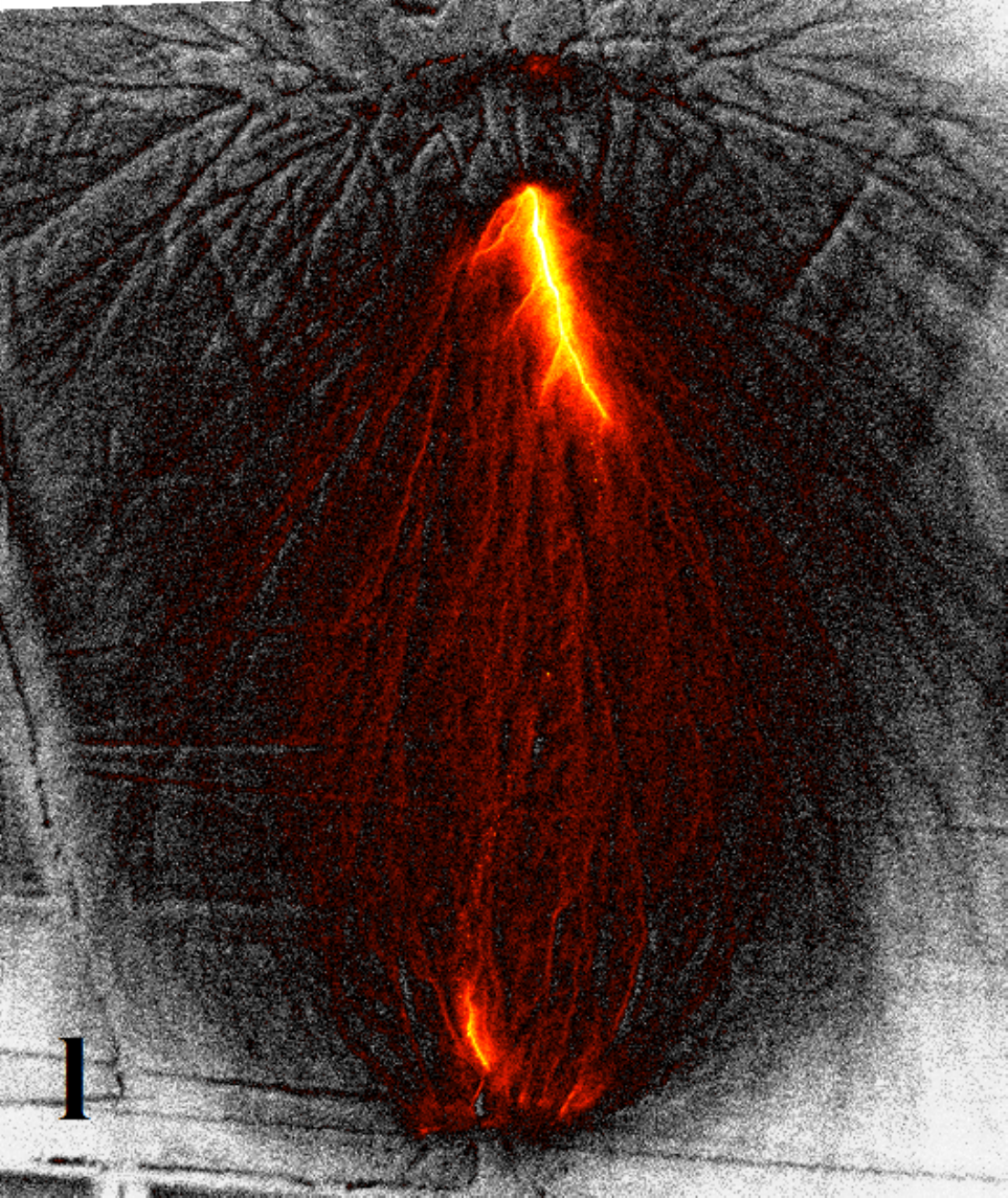}\end{minipage}
\begin{minipage}[b]{0.18\linewidth}
\includegraphics[width=30mm]{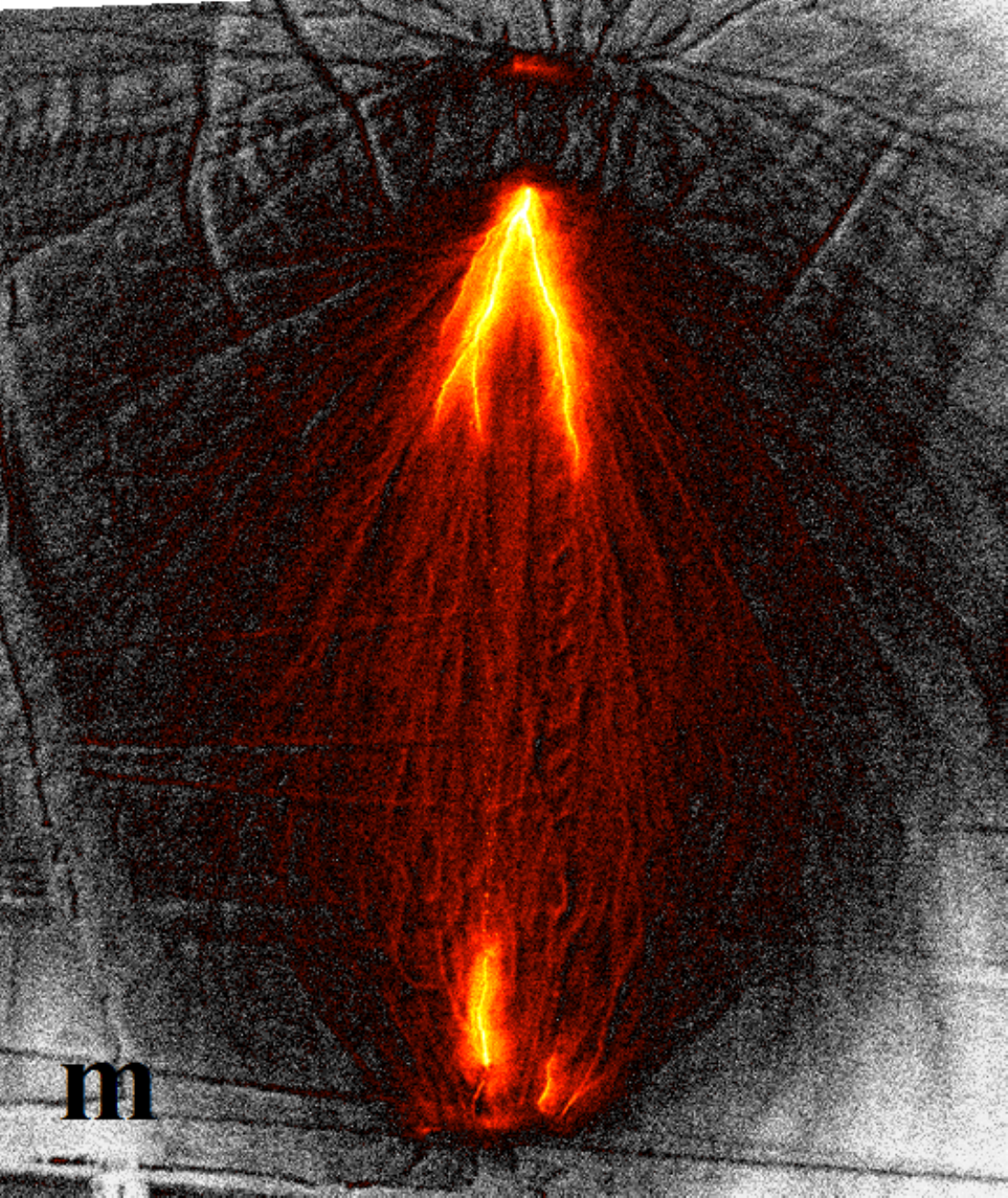}\end{minipage}
\begin{minipage}[b]{0.18\linewidth}
\includegraphics[width=30mm]{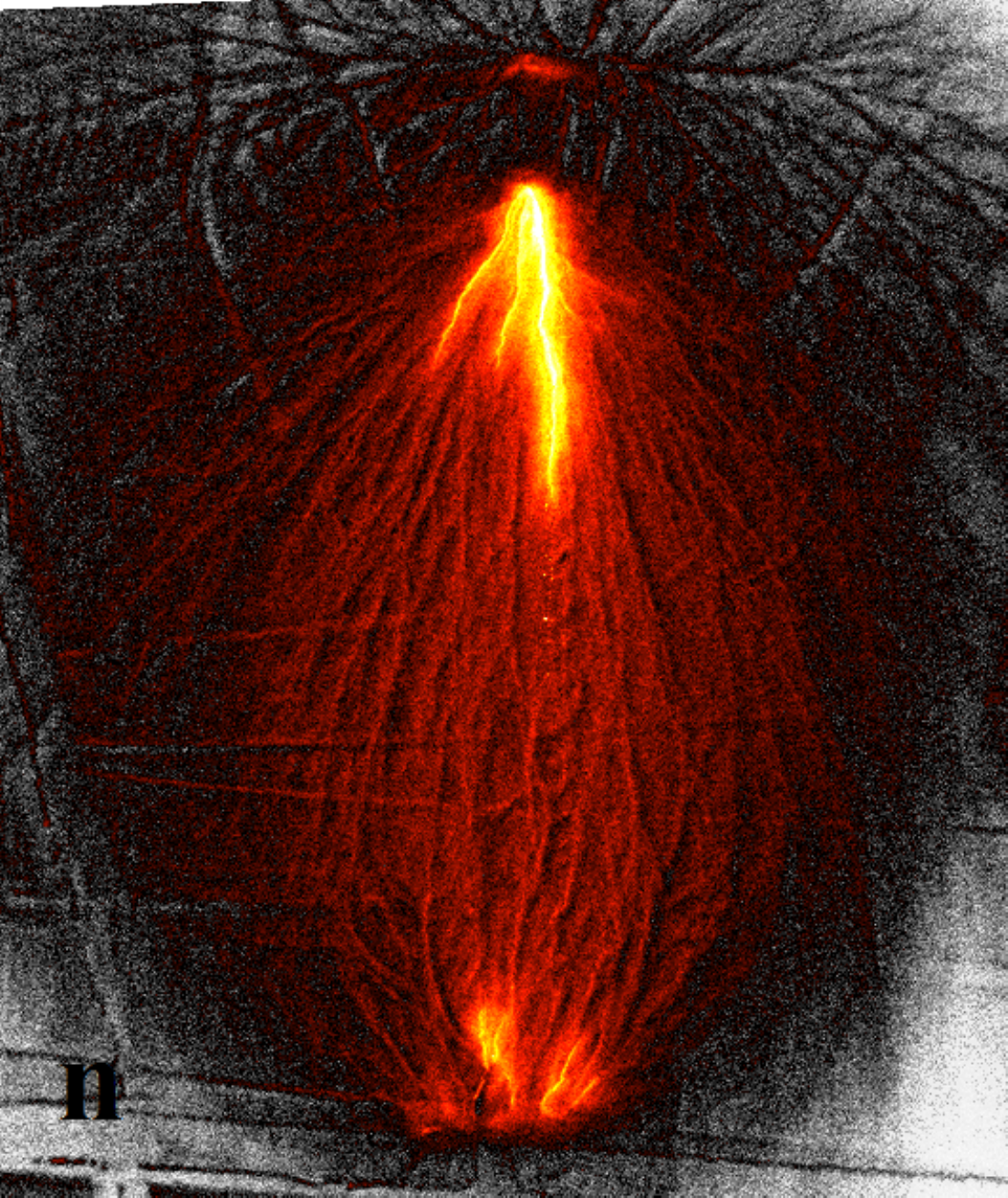}\end{minipage}
\begin{minipage}[b]{0.18\linewidth}
\includegraphics[width=30mm]{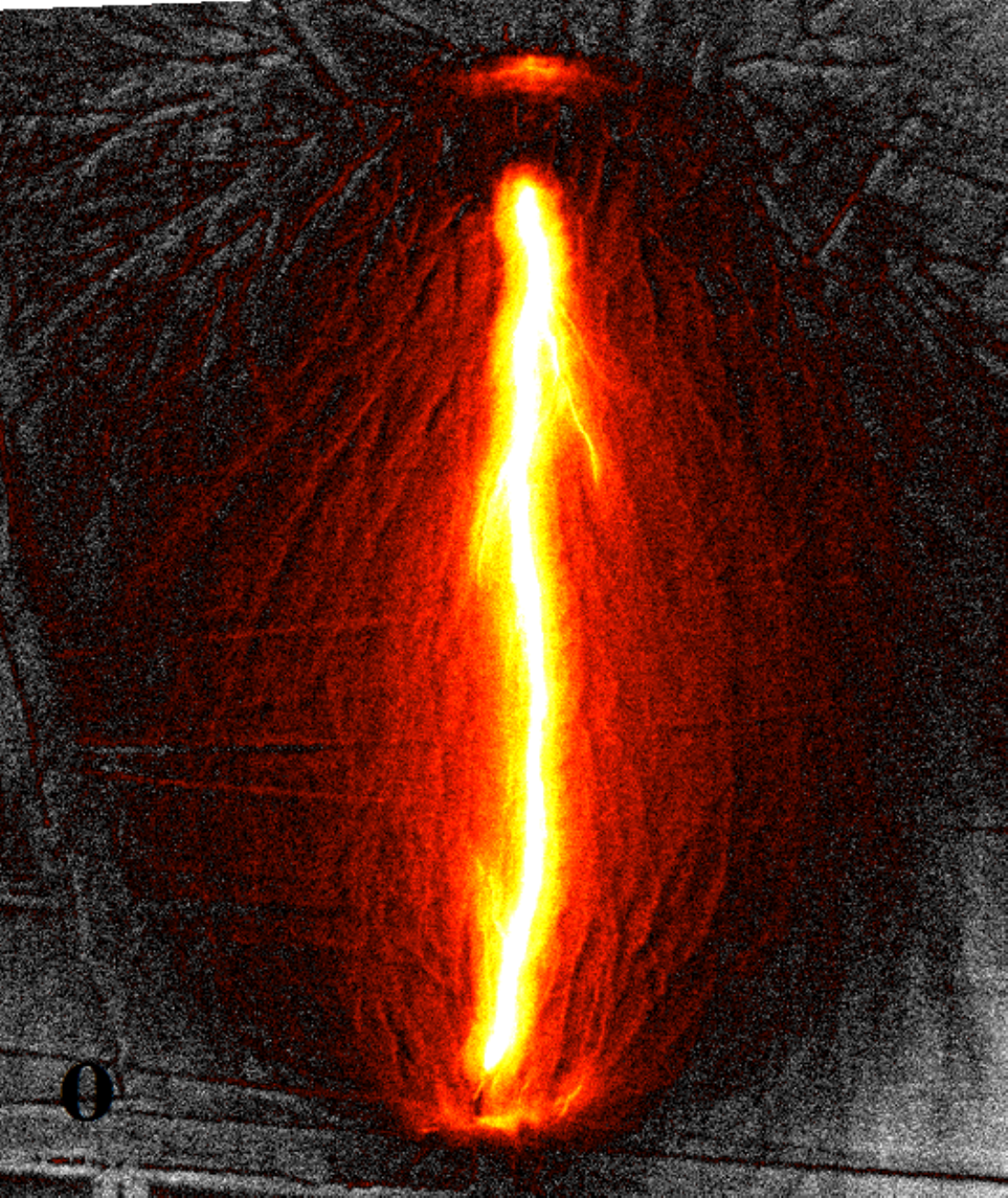}\end{minipage}
\begin{minipage}[b]{\linewidth}
\includegraphics[width=158mm, height=60mm]{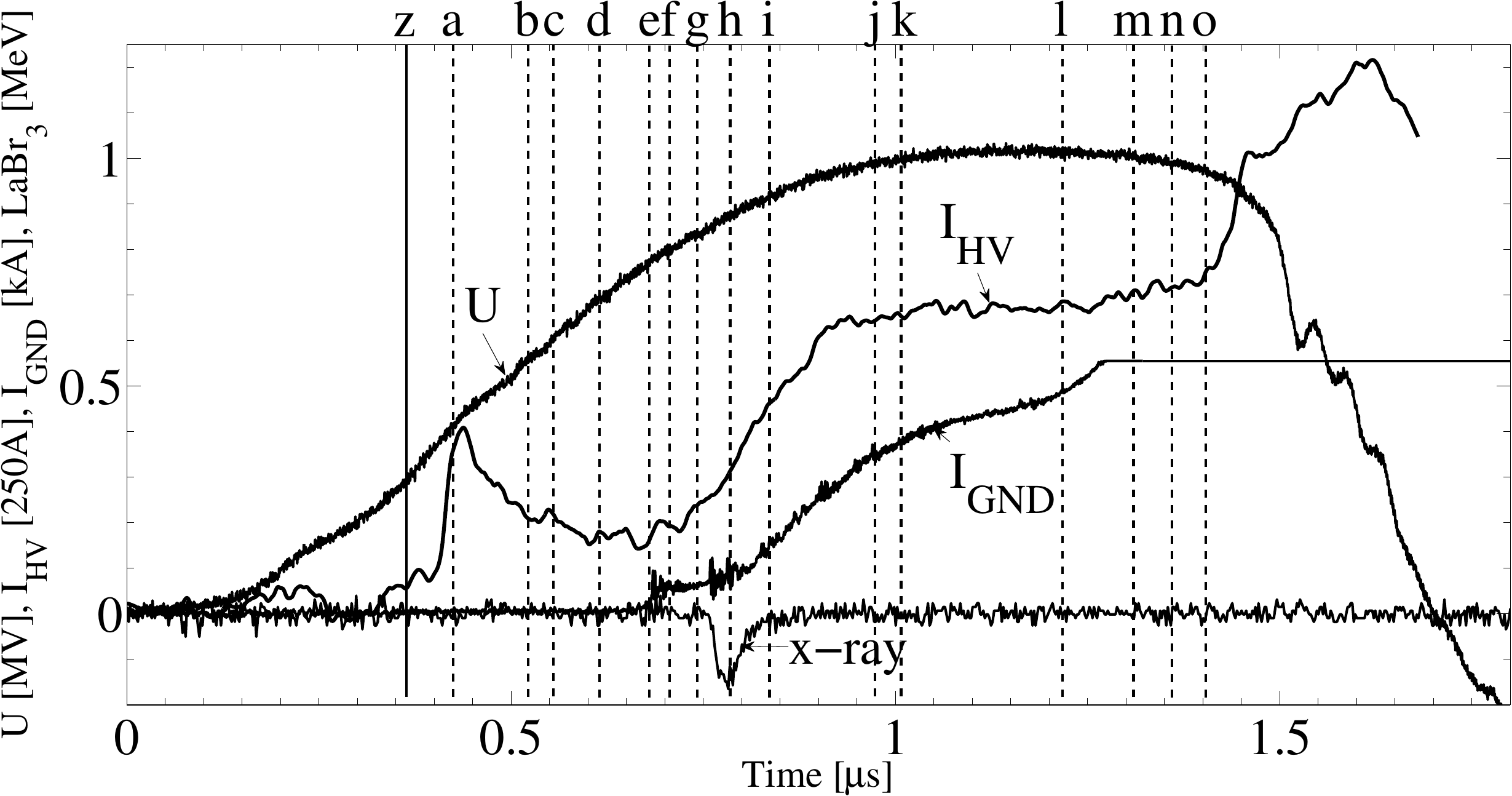}
\caption{Detailed development of one meter length discharge in consecutive time steps. Each picture corresponds to a single discharge. The shutter always opens at \textit{t}~=~0.36~$\mu$s (solid line \textit{z}). The exposure time varies from 60 (picture \textit{a}) to 1000 (picture \textit{o}) nanoseconds. The jitter with respect to the generator voltage and current is of the order of 0.04~$\mu$s. X-rays, voltage, cathode and anode currents are represented in the bottom plot. The data are those of picture \textit{l}. The moment of X-rays detection coincides with the connection between downward positive streamers and upward negative counter-streamers as shown in pictures \textit{e} -- \textit{h}.}
\label{fig:plot}
\end{minipage}
\end{figure}

The growth of the discharge is shown in \Fref{fig:plot}. Every picture corresponds to a single discharge. We applied a special linear colour coding scheme to the CCD output in order to enhance the faint streamers (pictures \textit{a} up to \textit{i}). For low luminosity the coding approaches an emulsion film negative with white as lowest light level. For the very bright leaders, this coding uses colour (pictures \textit{j} up to \textit{o}) up to the white level for maximum brightness. The solid vertical line \textit{z} at \textit{t}~=~0.36~$ \mu $s in the bottom panel corresponds to the shutter opening time. Dotted vertical lines \textit{a - o} correspond to the shutter closing times for the corresponding pictures. Thus, a single picture is time-integrated from the starting time \textit{z} until a camera closing time between 60 (picture \textit{a}) to 1000 (picture \textit{o}) nanoseconds later. Fast moving luminous streamers heads appear as streaks whose lengths correspond to their propagation lengths within the exposure time, similarly as in Figure 1 of \cite{Ebert2006}. The electrical signals corresponding to picture \textit{l} are represented at the bottom plot: the voltage waveform U, the currents at the high-voltage and at the grounded electrode, and the X-ray signal. On the time axis \textit{t}~=~0~$ \mu $s corresponds to the start of the voltage waveform. The pictures of the discharge development show a large similarity from discharge to discharge. There is also no significant difference in electrical characteristics. It means that one record represents most of the important steps in the process and can be used to describe the phenomena.

At the beginning, current up to 15~A  are measured on the HV electrode during 0.15 to 0.30 $ \mu $s, while the voltage starts to exceed 100~kV. The capacitive current due to the increasing voltage is much smaller. Therefore this small current leap must correspond to the first positive corona formation around the sharp tip of the HV electrode. This is probably a highly conductive inception cloud around the tip, similarly to the ones seen at lower voltages in \cite{Briels2008a,Briels2008b,Nijdam2011}. It extends radially until the electric field on its surface equals the breakdown field of 30~kV/cm. Image \textit{a} of \Fref{fig:plot} shows a cloud of approximately 6~cm radius at some later stage. The required charge is then about 1~$ \mu $C, in agreement with the current measurement when integrated up to 0.3~$ \mu $s.

After the leap, the first picture \textit{a} represents processes in the gap between 0.351 and 0.420~$ \mu $s: the voltage rises approximately linearly in time while the HV current increases rapidly. This high-voltage current jump occurs because of the additional corona formation followed by streamer bursts from the edges of the protection aluminium dish. Below the corona ignition zone there is a diffuse structure originating from the electrode tip (marked $t$). We interpret this as re-excitation of the inception cloud by the current needed for the streamers emanating from the cloud. The current through the grounded electrode remains zero and there is no light emission from that area.

Images \textit{b}, \textit{c}, \textit{d} show that a large number of streamers propagate downward in a rather synchronous manner. Downward streamers from the electrode tip and streamers, directed sidewards from the dish propagate at a speed of 2$\cdot10^{6}$~m/s in our experiments. The diameter of a single streamer is of the order of 1~cm measured at FWHM level by technique described in \cite{Nijdam2010}. It is remarkable that HV current gradually decreases then.

Near \textit{t}~=~0.65~$ \mu $s, the downward positive streamers approach the GND electrode and increase the local electric field. At that time for images \textit{d} and \textit{e} the distance between the positive streamer tips and the grounded electrode cone is approximately 20~cm. The electric field in this region is now strong enough to initiate the formation of negative streamers out of the grounded electrode. The very first current of the order of 10~Amps is detected on the GND current probe at that moment.

The current through the grounded electrode rises rapidly due to the formation of negative primary upward streamers from the sharp tip. In analogy to the counter-leaders that rise from the ground towards approaching lightning leaders, we will call these negative streamers counter-streamers. All cathode current initiations and steps are accompanied by damped oscillations at frequencies of several hundred MHz, well above the specified bandwidth of the probe 70~MHz.

Immediately after that, at the moment just before \textit{f}, a positive streamer connects to a negative counter-streamer. At \textit{t}~=~0.67~$\mu$s the first channel is formed by the encounter of two streamers -- positive from HV and negative from GND electrode. This encounter is again accompanied by high-frequency oscillations at the cathode current signal and more importantly by X-ray emission. This is the beginning of a gradual increase of the HV current. A few more connections between positive and negative streamers will take place later up to moment \textit{l} when all connection processes end. All of them are accompanied by HF oscillations. We refrain to speculate about the origin of these oscillations until a new current probe with an extended bandwidth is available. It seems safe to presume that many of these encounters are accompanied by X-rays. Some of the X-rays are detected by our detectors as first, second, etc. peaks. As plotted in \Fref{fig:plot}, we detected X-rays emitted by one on those connection processes.

The HV current measurements after picture \textit{i}, or current values larger than 250~A should be regarded with caution since the protecting diodes of the optical transmission start to act. The current graph shows the behaviour of the current, whether staying constant or rising, but not the actual values.

At the moment between \textit{i} and \textit{j} a hot conductive channel -- the leader -- arises. We can distinguish the leader from the streamers by its higher luminosity. The leader channel is much brighter than the streamers. Here we should recall that the optical and electrical parameters of the camera remain the same during all pictures from \textit{a} till \textit{o}. The leader formation process is the joule heating of one of the conductive streamer channels formed before. The leader tip itself does not emit any new visible streamers during its propagation in our experiments.

On the pictures \textit{j} -- \textit{n} the development of the positive leader is clearly shown. The leader grows in a highly ionized medium and repeats one of the streamer's paths. The speed of the positive leader is 1.5$\cdot$10$^6$~m/s.

Briefly before the exposure time of image \textit{l} ends, an upward leader rises from the cathode towards the anode. The cathode current $I_{GND}$ increases rapidly. This new counter-leader will propagate upward until it connects with the downward leader.

Continuous Ohmic heating of the conductive streamer channels between anode and cathode leads to a well-conducting leader channel at \textit{t}~=~1.4~$ \mu $s. On image \textit{o} we see the final breakdown that is accompanied by strong current rises and voltage drop. The high-voltage current curve is limited by the diodes on the input to the optical transmission line. The GND current is limited by clipping of the oscilloscope channel amplifier. The total vertical length of the discharge on image \textit{o} is equal to 1~meter.

The streamer head propagates and thus the streamer channel must have some finite conductivity, in spite of the low light emission of that channel. According to current understanding \cite{Li2009, Ebert2006, Ebert1997} the density of free electrons is about 10$ ^{13} $ to 10$ ^{14} $~cm$ ^{-3} $ in the freshly formed streamer channel depending on the field enhancement at the streamer tip. This electron density determines the conductivity of the channel. The conductivity will vary in space and time, and clear predictions cannot be given. From the experimental side it is difficult to retrieve conductivity values because many parallel streamers contribute to the current. In \Fref{fig:plot} we observe a sudden increase in luminosity starting at picture \textit{j}. One of the streamers channels heats up critically and increases its conductance to further develop into a leader. Because of the size, our leader is not comparable to a fully developed lightning leader.

The horizontal dimension of the cloud formed collectively by the numerous streamers is much larger than was expected before and equals approximately 1.4~m. This means, that all equipment placed at shorter distances will emit or receive streamers at certain times and will affect the measurements \cite{Nguyen2008}. This is why we changed the setup and applied the Velostat foil, as mentioned in \Sref{sec:experimental_setup}.

Note that the HV electrode carries a current 0.67~$\mu$s before the GND electrode does. This is due to the Ramo-Shockley effect. During that time, the loop of the current that includes the HV electrode closes via the displacement current to the environment, most notably to the soil and to the conducting safety fence near the spark gap. Only a very small part of the displacement current $dD/dt$ is captured by the grounded electrode.

The images also show broad slanted streaks and horizontal lines that clearly do not originate from the electrodes. These artefacts are caused by  reflections of the light from spark gaps in the Marx generator on the plies in the black background foil. It is worth mentioning that our camera detected emissions not only from the gap and the electrodes. The full conductive 9-meter connection line from the top of the divider to the spark gap and the top of the Marx generator emits streamers. Anyway, this illumination is relatively small and does not affect the pictures but it should be taken into account in the interpretation of the relation of current and voltage.

\begin{figure*}[h]
\centering
\includegraphics[width=80mm]{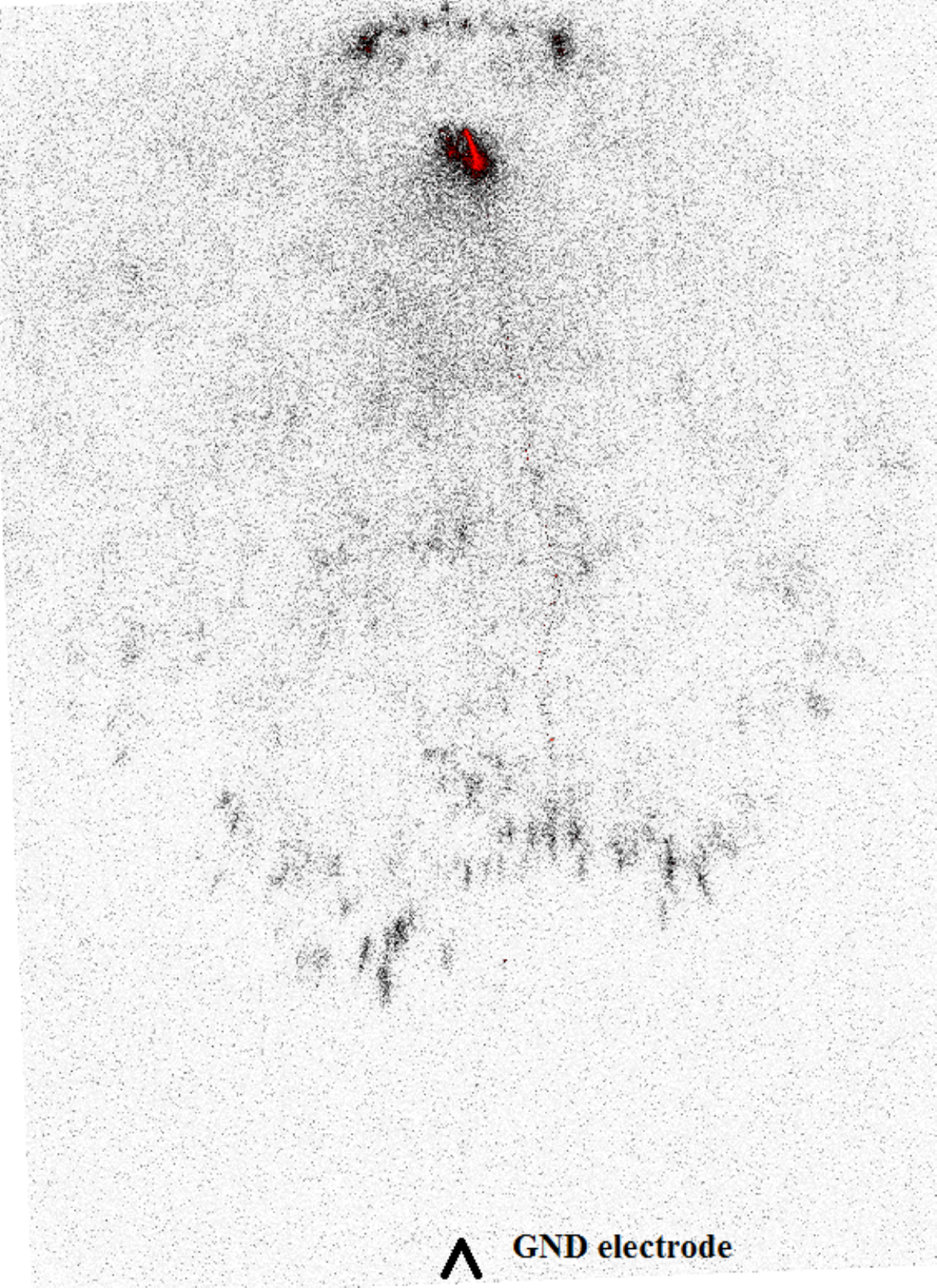}
\caption{Positive streamer heads approach the cathode. A short early positive leader exists before current and counter-streamers appear on the grounded electrode. The gap distance is 1.5~m, and the exposure time is 20~ns.}
\label{fig:1_5_meter_gap}
\end{figure*}

From the apparently constant luminosity throughout the length of streamer, we conclude that the positive streamer heads are "disconnected" from the HV electrode and fly like bullets through the medium, as also seen for streamers at lower voltages. Pictures with shorter exposure time (such as 20~ns in \Fref{fig:1_5_meter_gap}) confirm this. In the pictures shown in \Fref{fig:plot} we see the traces of those bullets. It is well known from streamer studies at lower voltages, that only the active ionization zones at the streamer heads emit light. At the streamer head the electric field is so high that electrons gain high energies and efficiently ionize and excite neutral molecules. The de-excitation of a certain nitrogen level (within about 1~ns in STP air) creates the light observed by the camera. Therefore the conducting channels stay invisible while the growing streamer tips are seen as "glowing dots". This was first photographed in  \cite{Blom1997} and is illustrated in Figure~1 of \cite{Ebert2006}. The same glowing dots at the tips of growing streamer channels have also been observed in sprite discharges high above thunderstorms at air pressures of about 10~microbar \cite{McHarg2007}, and a similar phenomenon also occurs as so-called "plasma bullets" in the complex gas mixture of pulsed atmospheric pressure plasma jets \cite{Sarron2011, Robert2012}.

In \Fref{fig:plot} the leader develops only after streamers and counter-streamers have connected. There is no leader observed with this 1~m gap before the first conductive channel is formed. But this is only a matter of scale. At larger gap lengths we did observe a positive leader before the positive streamers reached the cathode; this is illustrated in \Fref{fig:1_5_meter_gap} for a gap of 1.5 m.

In \Fref{fig:1_5_meter_gap} a faint speckle trace from the anode to cathode can be observed. This is caused by the huge brightness of the main discharge after the camera shutter was closed. The shutter is electronic and very good but not perfect. Some light leaks through the image intensifier even when it is off. But this fact gives us the opportunity to see the final breakdown path even on pre-breakdown pictures. That is why we decided not to suppress it further. For instance, in \Fref{fig:1_5_meter_gap} it is clear to see that breakdown followed the positive leader path. In \Fref{fig:plot} this trace also exists but due to longer exposure time and smaller scale it is almost indistinguishable.

\subsection{X-ray timing}
\label{sec:xray_timing}

\begin{figure*}[hbtp]
\centering
\includegraphics[width=80mm]{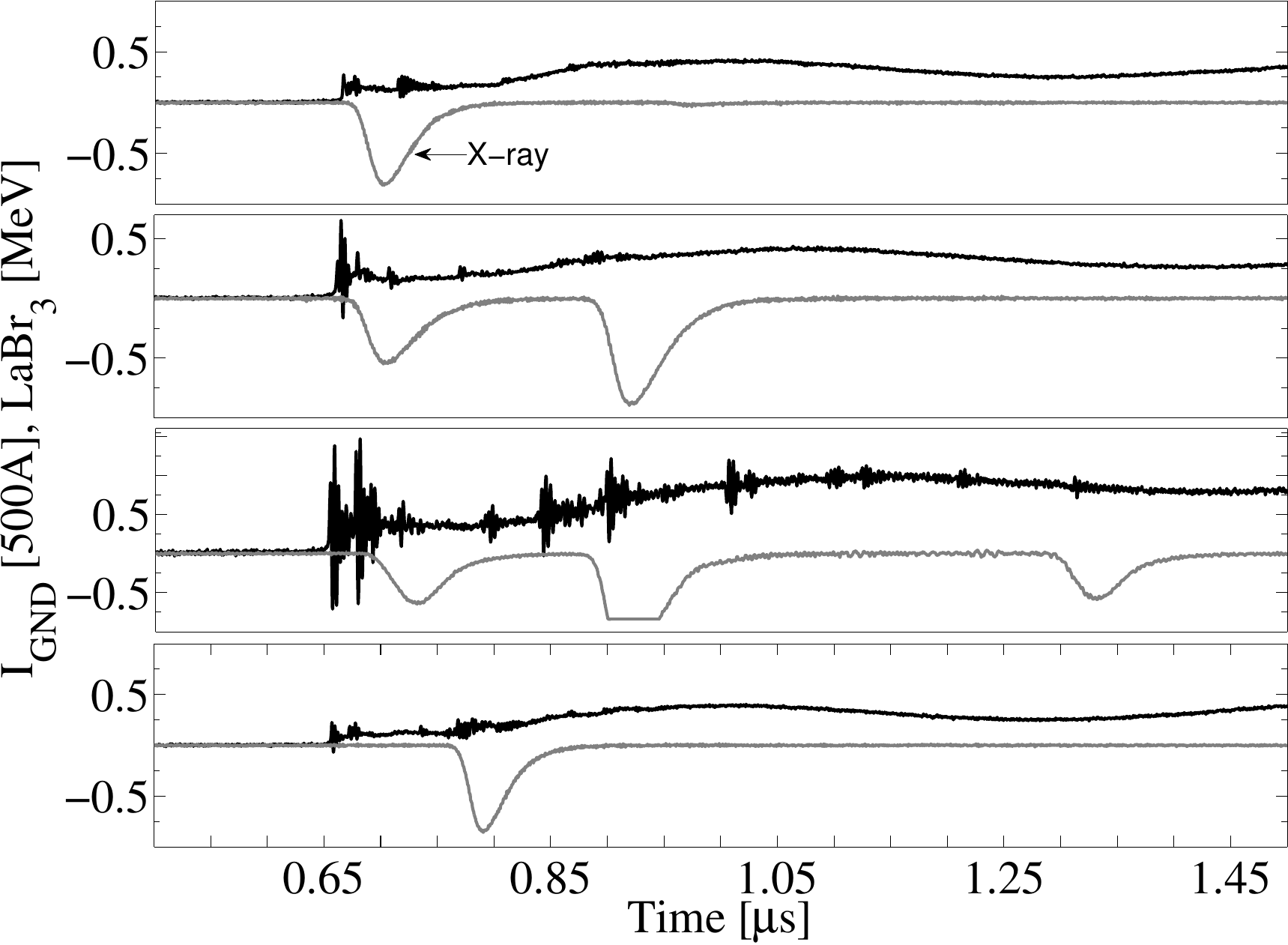}
\caption{Oscillations of the cathode current (black line) and correlated X-rays (grey line). All X-rays were accompanied by high-frequency oscillations of the cathode current that correspond to moments when positive and negative counter-streamers connect.}
\label{fig:4plots}
\end{figure*}

Some of the discharges had up to three X-rays peaks; see \Fref{fig:4plots}. Triple events were quite rare, of the order of 0.5\% occurrence. The discharges with two X-ray events occurred in up to 35\% of the discharge.

\begin{figure*}[hbtp]
\centering
\includegraphics[width=80mm]{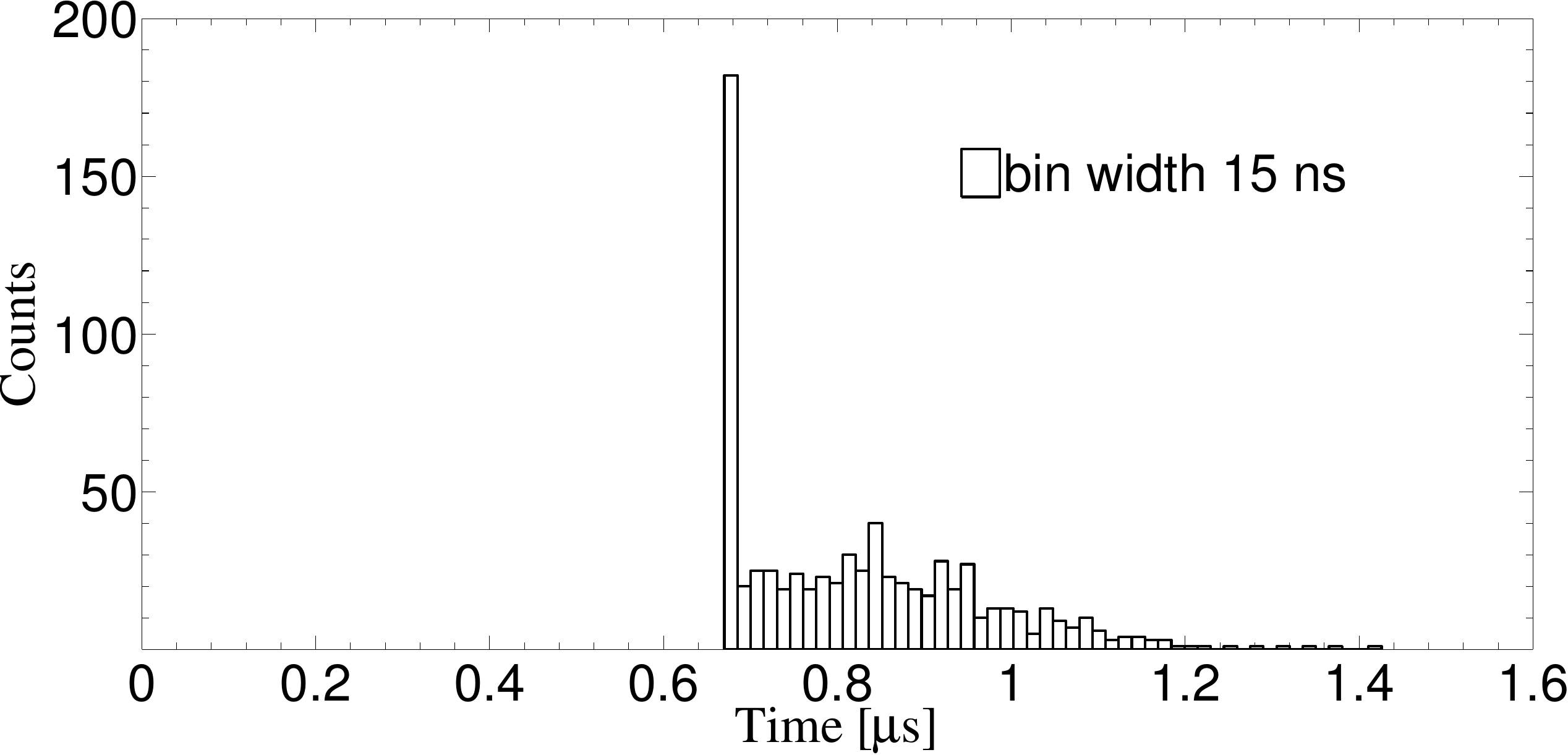}
\caption{X-rays time distribution. Timescale same as in \Fref{fig:plot} and \Fref{fig:4plots}. No X-rays were ever detected before cathode current started.}
\label{fig:time_hist}
\end{figure*}

The first X-ray photons were detected immediately after the first cathode current oscillation. \Fref{fig:time_hist} shows the distribution of X-ray occurrence times.

The leftmost peak coincides with the onset of the GND current. These immediate X-rays (immediate component) stem from the first early period of active streamer connection. Many upward and downward streamers meet at this time interval. The first encounters occur between positive downward streamers and negative upward counter-streamers emitted from the cone tip of the cathode. As will be shown later, the corresponding X-rays are the most energetic. Negative streamers emitted from the edge of the protection dish are involved in subsequent collisions. Subsequent X-rays still correlate in time with cathode oscillations and are produced in subsequent connection events. The number of these connections is higher than those of the first clashes, but the X-rays registered later have softer spectra. On the other hand, the HF oscillations amplitude is significantly larger during the first connections than during the subsequent ones. That means that the first connection events are more energetic and X-ray
productive events than those occurring later.

The total X-ray duration is of the order of 1~$\mu$s in our setup. The duration of one X-ray burst is determined by streamer collisions and can be less than the 11~ns of detector resolution.

We finally note that the background of the detector set at a discrimination level of 30~keV is approximately 50 counts per second and was determined by the measuring time needed to register 5000 detector pulses. This includes the contributions from cosmic rays, the laboratory environment and the internal isotope decay. The background remains constant during the day. The chance to observe any background signal that might mimic an X-ray signal from the gap during the X-ray duration time is less than 10$^{-4}$.

\subsection{Probability of X-ray registration}

We did not register X-rays in all discharges. The registration rate depends on the detector position as well. We determined the X-ray occurrence for different detector positions. \Tref{tab:occurrence} shows the result as the ratio of discharges with X-ray detection over the total number of discharges. To measure the registration rate at the points A and B we have placed a small EMC cabinet under the grounded electrode (as shown in \Fref{fig:setup}) and this cabinet remained there during all measurements in Series I and II. This cabinet also enhanced the X-ray production by streamers from the four sharp corners, from the surface of the cabinet and from the additional length of shielded cable to the current probe. For instance, at 2.1 meter distance from the gap, the X-ray occurrence was of the order of 7\% without the cabinet and 34\% with cabinet under otherwise the same conditions and effective registration areas.

\begin{table}
\caption{Occurrence of X-ray detection in different positions.}
\label{tab:occurrence}
\begin{indented}
\item[]
\begin{tabular}{@{}llll}
\ns \br
Point&Coordinates&Occurrence&Occurrence \\
&x;y (m)&out of surges&P (\%) \\
\mr
A$^1$ & 0.15;-0.13 & 290/441 & 66  \\

B$^1$ & 0.35;-0.13 & 51/120 & 43  \\

C$^1$ & 2.10;\- 0.15 & 54/160 & 34  \\

D$^1$ & 1.50;\- 0.15 & 97/140 & 69  \\

E$^2$ & 1.15;-0.3 & 10/10 & 100  \\

F$^2$ & 1.50;\- 0.6 & 29/30 & 97  \\

F' up$^2$ & 1.50;\- 0.6 & 19/50 & 38  \\

F' mid$^2$ & 1.50;\- 0.6 & 5/20 & 25  \\

F' down$^2$ & 1.50;\- 0.6 & 32/50 & 64  \\

G$^2$ & 1.50;\- 2.0 & 48/50 & 96  \\
\br
\end{tabular}
\item[] $^1$ Series I.
\item[] $^2$ Series II.
\end{indented}
\end{table}

The occurrence data for positions A,B,C,D were obtained in one series of experiments. E, F and G were obtained in another series two month later. All numerical values are shown in \Tref{tab:occurrence}.

It is still difficult to find the correlation between registration rate and detector distance due to the significant impact of the registration equipment at distances shorter than 1.5~m from the discharge (see \Fref{fig:plot}). But the occurrence of registration reduces with larger distance from the gap. In our opinion, the X-ray generation area can be approximated by a point-like source for larger distances. If we consider only geometrical decay (without absorption and scattering), the registration rate will follow an inverse square law for distances larger than 1.5~m, as suggested by the ratio of occurrence for positions D (69\%) and C (34\%). The run-away electrons from one negative streamer are certainly beamed. But for electron energies below 0.5~MeV, the Bremsstrahlung is rather isotropic (see Figure 3 in \cite{Chris2012}), in contrast to Bremsstrahlung from electrons with relativistic energies.

\subsection{Cathode shape}

\begin{figure*}[hbtp]
\centering
\includegraphics[width=80mm]{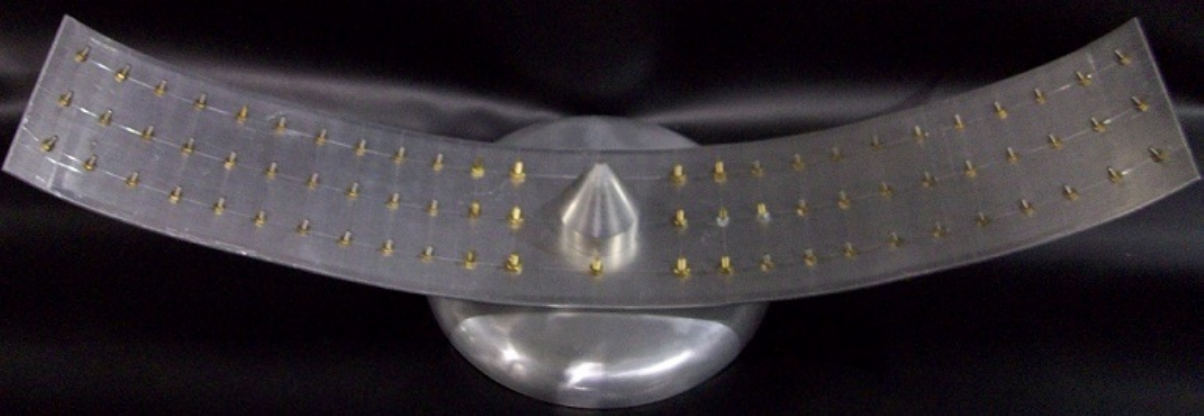}
\caption{Our alternative grounded electrode of 55 by 10~cm with 75 sharp points to generate upward negative streamers.}
\label{fig:belt}
\end{figure*}

Following the thought that mainly the upward negative streamers are responsible for the high-energetic electrons that generate the X-rays, the shape of GND electrode should significantly affect the number of negative streamers and thereby the X-ray occurrence. Implicitly we assume that the number of the positive streamers is constant and independent of the shape of the grounded electrode, at least during the first stages of discharge development. To check this, we made an aluminium belt of 55 by 10~cm with 75 sharp dots on its surface (\Fref{fig:belt}). We bent it according to the positive corona face as shown in picture \textit{d} of \Fref{fig:plot} and mounted it on the GND electrode. The small EMC cabinet was replaced by a post of the same height.

\begin{figure*}[hbtp]
\centering
\includegraphics[width=80mm]{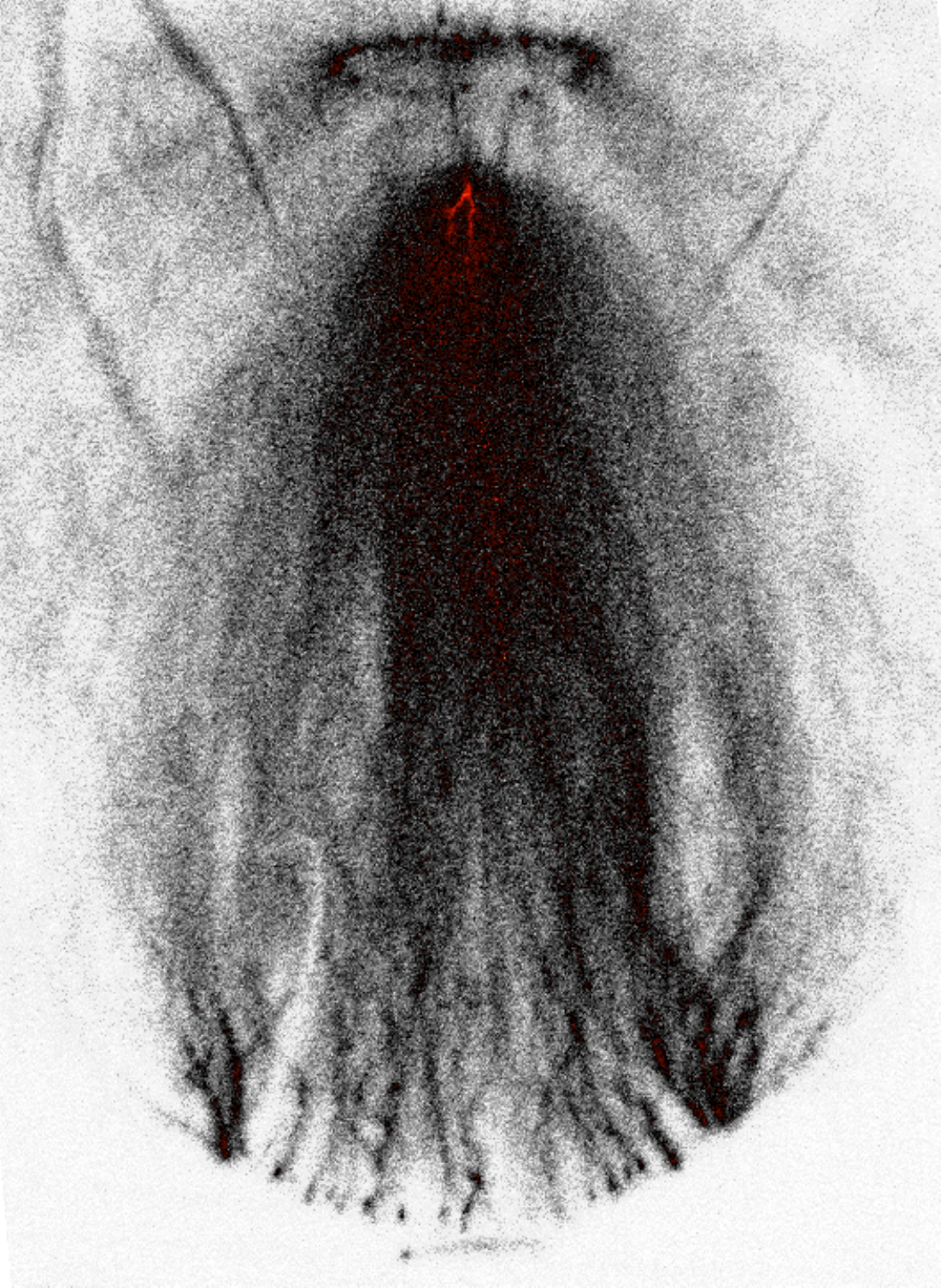}
\caption{Connection between positive downward and negative upward streamers initiated from the sharp belt. Exposure time is 300 ns; the gap length is 1 meter.}
\label{fig:belt_in_work}
\end{figure*}

\Fref{fig:belt_in_work} shows that the total number of negative streamers is greatly enhanced. The effective connection area and the number of encounters between positive streamers and negative counter-streamers significantly increased while the shape of the positive corona and the streamer density remain the same.

The X-rays were observed from 2.1~m distance by two LaBr$_{3}$ detectors. The total effective scintillator area was 23~cm$^{2}$. The X-ray occurrence increased from 7\% without the belt up to 70\% with the belt. The number of cathode current oscillations also remarkably increased. Without belt we registered triple X-ray signals in 0.5\% of the discharges and with belt in 15\%. So, a fakir’s bed strongly enhances the X-ray production. The small EMC cabinet under the cathode (row C in \Tref{tab:occurrence}) caused a similar increase. This cabinet causes impact on all points in the table. So the all X-ray data during Series I and Series II are intensified by additional negative counter-streamers from the small EMC cabinet.

\subsection{The region of X-ray emission}
\label{sec:the_region_of_xray_emission}

In order to determine where the X-rays are generated, we limited the field of view of the detector by lead tubes of 1 and 2.5~cm thickness covering the detector over its full length. This thickness is sufficient as will be shown in \Sref{sec:energy_spectra} on attenuation curves. A number of 100 surges were enough to determine the probability of seeing X-rays. But we refrained from an analysis of the X-ray energy spectra in these experiments. We pointed the detectors to different parts of the gap: the HV electrode, the GND electrode and the middle (see cones in \Fref{fig:setup}, point F'). With the detector pointing to the cathode (F' down, \Tref{tab:occurrence}), two times more events were observed than with the detector pointing to the anode (F' up, \Tref{tab:occurrence}). When we directed the detector to the middle position, we found less X-ray events (F' mid, in \Tref{tab:occurrence}) than from the anode and cathode region.

Both first and subsequent X-rays were registered to come from the cathode, anode and central regions. In some cases X-rays from the cathode and from the anode region were detected simultaneously, to within the time resolution of our detectors. It means that the high-energy electrons can emerge from the region between the streamers (see images \textit{e}, \textit{f}, \textit{g} in \Fref{fig:plot}) but create Bremsstrahlung over the full length of the gap including the anode and the surrounding area.The time that relativistic electrons need to cross the full gap is of the order of 3~nanoseconds or less, which is within the time resolution of the X-ray detectors.

\subsection{Energy spectra and attenuation curves}
\label{sec:energy_spectra}

\begin{figure*}[hbtp]
\centering
\includegraphics[width=80mm]{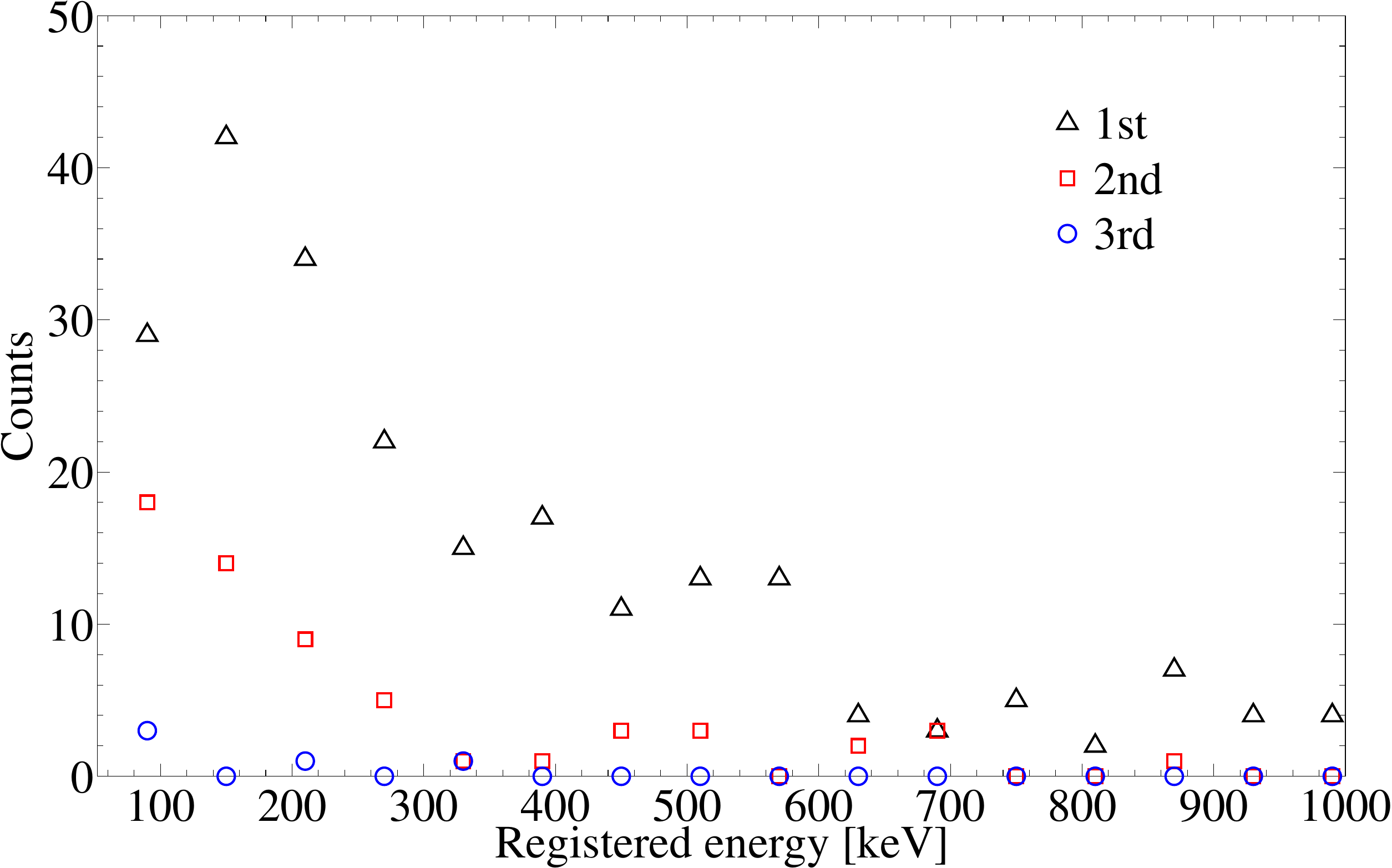}
\caption{Energy spectra of the detected X-ray photons. The three sets of data \opentriangle, \opensquare  and \opencircle correspond to the first, second and third x-ray burst respectively.}
\label{fig:pos_spectra}
\end{figure*}

\begin{figure*}[hbtp]
\centering
\includegraphics[width=80mm]{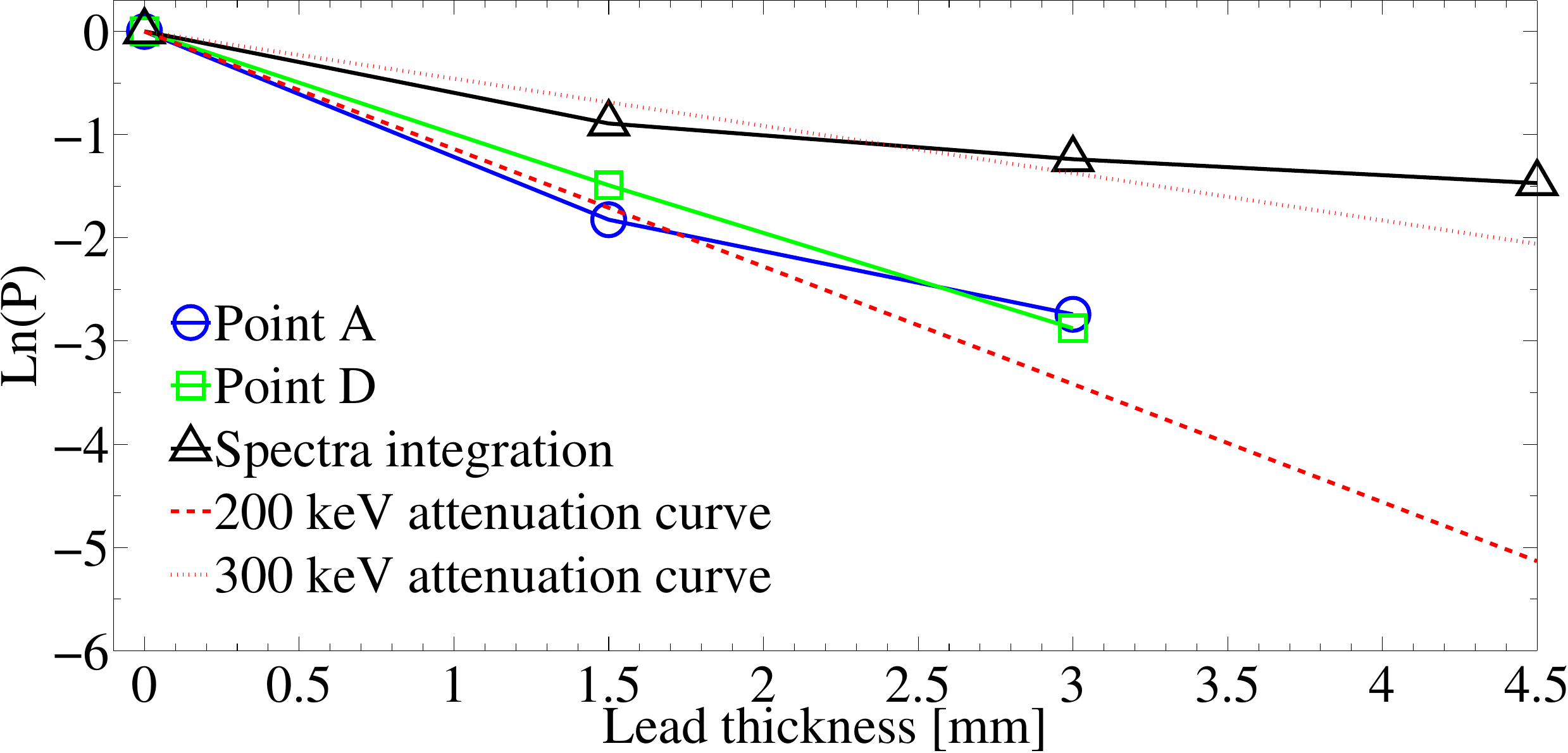}
\caption{X-ray lead attenuation curves at point A close to the grounded electrode and at point D at 1.5~m from the gap (\Fref{fig:setup}). The integrated spectrum $\opentriangle$ and attenuation lines for 200 and 300 keV from the NIST database are shown as well. The connecting lines between the measurement points are drawn as a guide to the eye.}
\label{fig:attenuation}
\end{figure*}

We collected data of 441 discharges with the detector at position A (\Fref{fig:setup}) inside the small EMC cabinet. The number of occurrences as a function of energy is shown in \Fref{fig:pos_spectra}, where each X-ray signal in a discharge is counted individually. The energy scale is divided in bins of 60~keV. Please note that the first X-ray signal can be delayed with respect to the first cathode current. This is because we probably did not register X-rays produced by first streamer encounters. An example is shown in the two lower records of \Fref{fig:4plots}. X-ray in bottom plot is marked as first X-ray. The first occurrence is most often the most energetic. Even a single event with up to 3.4~MeV has been observed. For the sake of clarity we limited the energy scale to 1~MeV, which is also the maximum energy a single electron can acquire with the available voltage. The later X-ray signals are less frequent by a factor of three or more. But the energy of the second signal can still be comparable to or larger than the first; see for example the third record in \Fref{fig:4plots}. As mentioned in \Sref{sec:xray_timing}, the X-ray signals correlate well in time with the connection moments of positive and negative streamers as shown in the pictures of \Fref{fig:plot}. In order to determine the actual photon energy of the observed X-rays we carried out a series of experiments with lead attenuators in front of the collimated detector. One detector was again mounted in the small cabinet at position A close to the gap, another in the large cabinet at 1.5~m from the gap. Caps of various thicknesses were used: 1.5, 3.0, 4.5~mm, and 50 to 100 discharges were investigated for every cap thickness. The green ($\opensquare$) and blue ($\opencircle$) data in \Fref{fig:attenuation} show the dependence of X-ray occurrence on the lead thickness.

\Fref{fig:attenuation} also shows the absorption curve ($\opentriangle$) that would be observed assuming that the amplitude of the X-ray signals without absorber would one-to-one correspond to single photon energy. This spectrum integrated absorption curve lies definitely above the experimental data. This indicates that many of the high energy signals are due to pile up: several lesser energy X-ray photons are registered simultaneously within the 11~ns time resolution of the detector. Two additional lines represent the lead absorption for monoenergetic photons of 200 and 300~keV, calculated with data from \cite{Berger1998}. The experimental data fit best to the 200~keV line.
However, with the wide spectrum of X-rays, one may question whether it is allowed to convert the attenuation curve into an equivalent energy.

\begin{figure*}[hbtp]
\centering
\includegraphics[width=80mm]{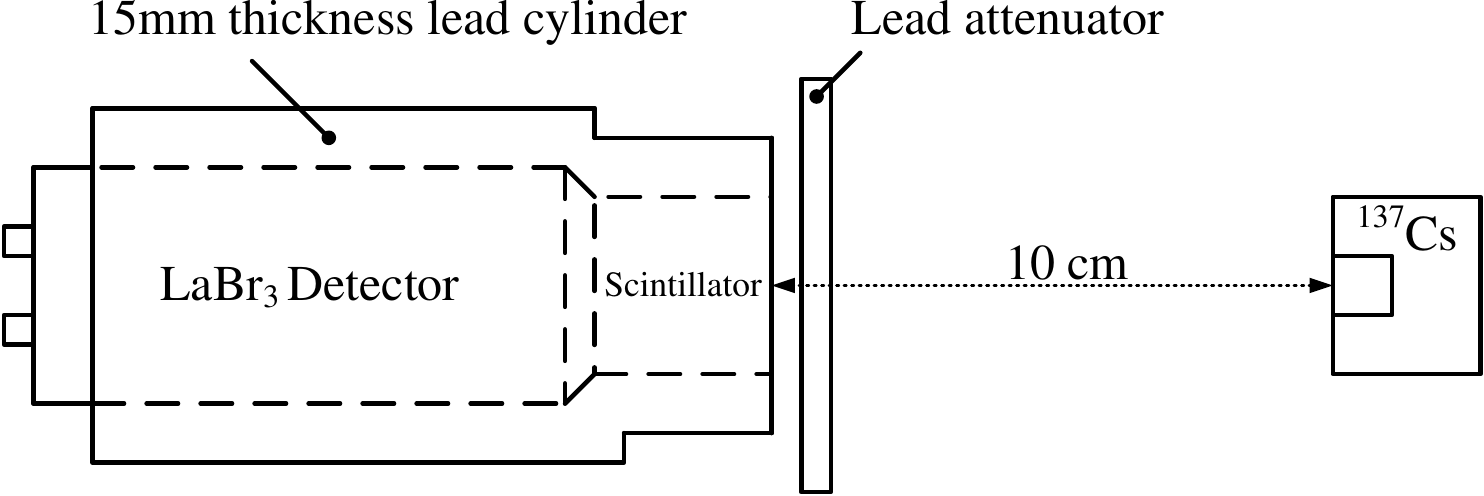}
\caption{Lead attenuation setup of 662~keV photons from $^{137}$Cs radioactive source.}
\label{fig:cs_setup}
\end{figure*}

\begin{figure*}[hbtp]
\centering
\includegraphics[width=80mm]{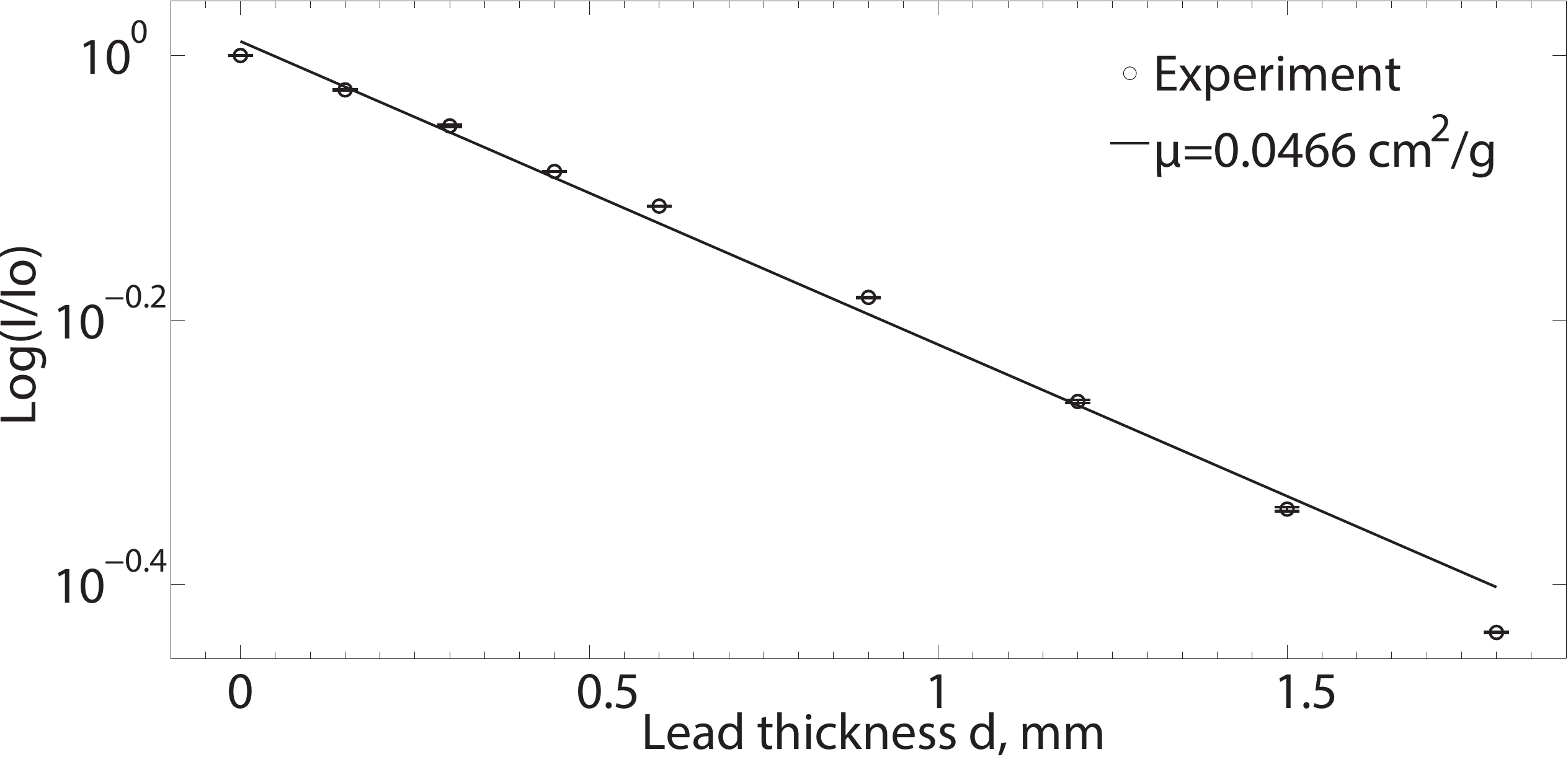}
\caption{Attenuation of 662 keV gammas by lead and comparison with NIST data.}
\label{fig:cs_attenuation}
\end{figure*}

In order to get some insight into the energy spectrum, we obtained the attenuation curves of 662~keV gammas from a near to monochromatic $^{137}$Cs source in a similar setup (\Fref{fig:cs_setup}), where we focused on the attenuation of all photons arriving at the detector. The question is whether it can be expected that the attenuation curve is a straight line on a semi-logarithmic plot? The experimental linear attenuation coefficient (curve slope) is $\mu~=~4.66\cdot10 ^{-2}$~cm$^{2}$/g. The coefficient $\mu$ consist of two parts: $\tau$ -- the photoelectric component and $\varepsilon$ -- the Compton effect component. Pair production in the lead can be neglected due to the energy threshold of 1.022~MeV. To draw the attenuation curve (\Fref{fig:cs_attenuation}), we counted all photons that penetrated through the lead attenuator. We did not select on energy. Partial absorption by the lead is then not accounted for. This also occurs with the X-rays from the spark, where the photon energy before attenuation is a priori unknown.

According to the NIST database \cite{Berger1998}, the photoelectric absorption coefficient for 662~keV photons equals $\tau~=~4.33\cdot10^{-2}$~cm$^2$/g. The small difference between the total attenuation coefficient $\mu$ and the photoelectric absorption coefficient from the NIST data implies that the Compton effect inside the lead for this setup is negligible. Most of the Compton-scattered photons penetrate the lead attenuator and hit the scintillator. If this statement is correct for 662~keV energy photons (Compton effect comparable to photoelectric absorption) it is particularly correct for 200~keV energy photons, where the photoelectric component is 10 times larger than the Compton component. Consequently, we should only take the photoelectric part into account for the energy reconstruction from attenuation curves.

\section{Discussion and conclusions}
\label{sec:discussion}

Our experimental results clearly demonstrate that the presence of negative streamers is a necessary condition for X-ray generation in meter long positive discharges in the laboratory. There were no X-rays observed before the negative counter-streamers initiated from the grounded electrode. Most of the detected X-rays correlated with high-frequency current oscillations on the current probe of the grounded electrode. By increasing the number of negative streamers with our fakir’s bed electrode we dramatically increased the occurrence of X-rays. That negative streamers play a role in X-ray generation is consistent with the theoretical papers quoted in the introduction \cite{Moss2006,Li2009,Chanrion2010,Celestin2011} that show how electrons are accelerated out of the streamer tip and away from the electrode. The Bremsstrahlung photons are therefore generated in open air, and not at some metal parts. The X-ray localization confirms this statement.

The current data together with the pictures show that the high-energy electrons necessary for the X-rays correlate in time with the encounter of positive downward and negative upward streamers. It is most probable that the run-away process occurs in the region between positive and negative streamer tips just before the connection because the electric field and total available voltage then and there are higher than anywhere else. It would be a hard but interesting task to assign an individual connection to the run-away process. As mentioned before, run-away process and X-ray generation occur within the 11 ns time resolution of our detectors, but do no need to coincide in space. Streamer encounters generate ns-fast X-ray bursts.

The absorption data are consistent with a predominant photon energy of 200~keV.

There is only a small allusion of non-homogeneity of the X-ray emission present in our data. The spectra in point G (\Fref{fig:setup}) seem to be more energetic than at other points. It is unlikely that X-rays were emitted by other parts of the Lightning Surge Generator installation due to absence of negative streamers. No strong orientation dependence of the X-ray emission was found in our experiments, in agreement with \cite{Chris2012}.

The pictures discriminate very well between the streamers and the leaders and clearly indicate the moment when the leader starts to develop. Streamers are less bright than leaders. Both appear to have nearly the same velocity (2$\cdot$10$^6$~m/s) and thickness of the order of 1~cm for the streamer channel and for the leader channel core.
All thicknesses are measured at FWHM level by the technique described in \cite{Nijdam2010}. Around the leader core there is a region of lower luminosity with a diameter of about 7~cm, that might be identified as leader sheath. A more detailed discussion requires knowledge of the contrast transfer function of the optical chain consisting of lens, image intensifier and CCD. This information is not available yet.

A full breakdown is not a necessary condition to generate X-rays because the X-rays are generated in the streamer phase. Even the leader is not necessary. But it is difficult to avoid full breakdown of the high voltage gap in our setup. The usual way is to short-circuit the generator by a second gap. This gap would then also generate X-rays.

It is interesting to speculate and to extrapolate our results to negative high-voltage discharges. There the space stem that can appear in front of the negative leader can play the role of the positive electrode.
The main point from our work is that negative streamers approaching positive ones appear to be a necessary condition for X-rays in our setup, even for positive high-voltage discharges. Though there are certainly features in our setup that are not representative for natural lightning like the gap length and the voltage amplitude and shape, we observe generic phenomena of streamer and leader formation, and our measurements show the dominant role of the negative streamer corona for the production of hard radiation; such negative streamer coronas appear as well at the negative ends of lightning leaders and space stems.

\ack{P.K. acknowledges financial support by STW-project~10757, where Stichting Technische Wetenschappen (STW) is part of The Netherlands' Organization for Scientific Research NWO. Dr. A.J.M. Pemen kindly provided the 4Picos camera used in the experiment.}
\section*{References}

\providecommand{\newblock}{}

\end{document}